\documentclass[preprint,12pt]{aastex}
\usepackage{epstopdf}
\usepackage{epsfig}
\usepackage{longtable,lscape}

\shorttitle{Near-IR WR Survey. II.}
\shortauthors{Shara et al}

\begin{document}


\title{A Near-Infrared Survey of the Inner Galactic Plane for Wolf-Rayet Stars II. Going Fainter: 71 More New WR Stars}


\author{Michael M. Shara\altaffilmark{1}}
\affil{American Museum of Natural History, 79th Street and Central Park West, New York, NY, 10024-5192}
\email{mshara@amnh.org}

\author{Jacqueline K. Faherty\altaffilmark{1}}
\affil{American Museum of Natural History, 79th Street and Central Park West, New York, NY, 10024-5192}
\email{jfaherty@amnh.org}

\author{David Zurek\altaffilmark{1}}
\affil{American Museum of Natural History, 79th Street and Central Park West, New York, NY, 10024-5192}
\email{dzurek@amnh.org}

\author{Anthony F. J. Moffat}
\affil{D\'epartement de Physique, Universit\'e de Montr\'eal, CP 6128 Succ. C-V, Montr\'eal, QC, H3C 3J7, Canada }
\email{moffat@astro.umontreal.ca }

\author{Jill Gerke}
\affil{Department of Astronomy, Ohio State University, Columbus OH 43210-1173}
\email{gerke@astronomy.ohio-state.edu}

\author{Ren\'e Doyon }
\affil{D\'epartement de Physique, Universit\'e de Montr\'eal, CP 6128, Succ. C-V, Montr\'eal, QC, H3C 3J7, Canada }
\email{doyon@astro.umontreal.ca }

\author{Etienne Artigau}
\affil{D\'epartement de Physique, Universit\'e Laval, Pavillon Vachon, Quebec City, QC, G1K 7P4 Canada}
\email{artigau@astro.umontreal.ca}

\author{Laurent Drissen }
\affil{D\'epartement de Physique, Universit\'e Laval, Pavillon Vachon, Quebec City, QC, G1K 7P4 Canada}
\email{ldrissen@phy.ulaval.ca}

\altaffiltext{1}{Visiting Astronomer at the Infrared Telescope Facility, which is operated by the University of Hawaii under Cooperative Agreement no. NNX-08AE38A with the National Aeronautics and Space Administration, Science Mission Directorate, Planetary Astronomy Program }

\begin{abstract}
We are continuing a J, K and narrow-band imaging survey of 300 square degrees of the plane of the Galaxy, searching for new Wolf-Rayet stars. Our survey spans 150 degrees in Galactic longitude and reaches 1 degree above and below the Galactic plane. The survey has a useful limiting magnitude of K = 15 over most of the observed Galactic plane, and K = 14 (due to severe crowding) within a few degrees of the Galactic center. Thousands of emission line candidates have been detected. In spectrographic follow-ups of 146 relatively bright WR star candidates we have re-examined 11 previously known WC and WN stars and discovered 71 new WR stars, 17 of type WN and 54 of type WC. Our latest image analysis pipeline now picks out WR stars with a 57\% success rate. Star subtype assignments have been confirmed with K band spectra, and distances approximated using the method of spectroscopic parallax. Some of the new WR stars are amongst the most distant known in our Galaxy. The distribution of these new WR stars is beginning to trace the locations of massive stars along the distant spiral arms of the Milky Way.

\end{abstract}

\section{Introduction and Motivation}

Most Population I Wolf-Rayet (WR) stars are the helium-burning descendants of the most massive stars (with initial masses greater than $\sim 20 M_\sun$ at $Z_{\odot}$). They are also amongst the most luminous stars known. Their powerful winds ($\dot{M} \sim 10^{-5} M_\sun yr^{-1}$ ) display strong, broad emission lines of helium, and either nitrogen (WN subtypes) or carbon/oxygen (WC/WO subtypes) - the defining observational characteristics of WR stars. Because of their relatively short lifetimes (about $5\times10^5$ years, which is roughly 10\% of the star's total lifetime), WR stars are excellent tracers of recent star formation. They are also believed to be type Ib or Ic supernova progenitors, because they have removed their outer H-rich layers (WN) or even He-rich layers (WC/WO) (but see also \citealt{sma09}).

Galactic distribution models predict that $\sim$1000-6500 WR stars are expected (\citealt{sha99}; \citealt{sha09}; \citealt{vdh01}) in total, but this assumes that massive stars are uniformly distributed throughout the Milky Way. If the total WR star population is as high as 6500 then one might erupt as a Type Ib or Ic supernova within a few generations, as each lasts $\sim5\times10^5$ yrs. The clear identification of a WR star as the progenitor of one of these eruptions would be a dramatic confirmation of a key prediction of stellar evolution theory. It would be no less valuable to show that a type Ib or Ic progenitor did NOT have a WR star progenitor.

The prediction of a second test of massive star evolution theory follows from the radial metalicity gradient across our Galaxy \citep{sma97}. Higher Z is predicted to lead to stronger stellar winds that reveal the deeper parts of massive stars more quickly. This suggests that the WC/WN number-ratio must increase sharply in the inner parts of the Milky Way relative to what we observe in the Solar neighborhood \citep{mey05}. This is consistent with what is presently observed \citep{sha99}, but  before we began the survey that is the subject of this paper only about 300 WR stars had been identified in the Milky Way \citep{vdh06}. Carrying out these important tests of stellar evolution theory demands a much larger and more complete census of Galactic Wolf-Rayet stars - particularly in regions at different Z than the solar neighborhood - than has hitherto been possible. Optical narrow-band surveys have been severely limited by interstellar extinction \citep{sha99}, so a majority of the known WR stars lie within a few kpc of the Sun. The only reasonable way to locate the vast majority of the Galactic WR stars is to search for them in the near infrared where the Milky Way is quite transparent.

In \citet{sha09} (hereafter Paper I) we described a new narrowband infrared imaging survey of much of the Galactic Plane. The goal outlined in that paper was to locate and characterize 90\% of the WR stars in the Galaxy within a decade. Details of the infrared camera, filters, telescope and image processing used to reduce the 77,000 science and dome flat images (taken in 2005 and 2006) are given in Paper I. We also described our candidate selection criteria, and we focused on 173 bright candidate targets with emission-band magnitudes brighter than K = 11.5 for followup spectroscopy. Our exploratory 2007 spectrographic run, detailed in Paper I, resulted in the confirmation of 41 new WR stars: 15 WN and 26 WC, represented a nearly 24\% success rate.

This paper reports the results of further followup spectroscopy (carried out in 2009) of 146 candidates brighter than K = 12.5 . We continued to use the aperture photometry methodology described in Paper I to compare the magnitudes of all the stars detected in both broadband and narrow-band (HeI, HeII, CIV and Brackett-gamma) images. We compared candidates' images carefully by eye to remove spurious or doubtful stars, and culled stars of lower statistical significance. This resulted in a significantly higher success rate than reported in Paper I. After this paper was completed, a complimentary effort to locate WR stars using NIR and mid-IR colors was published by \cite{mau11}. Using the technique first outlined by \citep{had07} they located 60 new WR stars. In Section 2 we describe our spectrographic observations and the data reduction procedure we used. We present the spectra and spectral types, and derive the distances and spatial distribution in the Galaxy of our new WR stars in Section 3. In Section 4 we briefly note a new ring nebula WR star and two WR stars in a compact cluster. In section 5 we discuss the completeness of this survey, and the complementarity of the narrowband and colors-based surveys. The finder charts of the new WR stars are presented in section 6, and we summarize our results in Section 7.

\section{Observations: Near-Infrared Spectroscopy with SPEX}
Near-IR spectra were obtained of 146 candidate WR stars with the SpeX spectrograph mounted on the 3m NASA Infrared Telescope Facility (IRTF) over eleven half nights in August 2009. The conditions of this run were excellent with average seeing (0.5 -0.8 $\arcsec$ at $\it{J}$).    We operated in cross-dispersed mode with the 0.5$\arcsec$ slit aligned and obtained an average resolving power of  $\lambda$/$\Delta$ $\lambda$ $\sim$1200. The near-infrared spectral data spanned 0.8 - 2.4 $\mu$m .  Each target was first acquired in the guider camera.  We evaluated each candidate after a single AB dither pattern with exposure times varying from 30s for our brightest targets to 200s for our faintest.  Once we had confirmed the presence of emission lines we began a second set of AB images so each WR star had four images obtained with an ABBA dither pattern along the slit.   To minimize slew and calibration target time we chose subsequent targets closeby in the sky.  An A0V star was observed after each several targets (typically 4-5) at a similar airmass for flux calibration and telluric correction.  Internal flat-field and Ar arc lamp exposures were also acquired for pixel response and wavelength calibration, respectively.  We also acquired a spectrum of almost all known spectral subtypes of Wolf-Rayet star. All data were reduced using SpeXtool version 3.3 \citep{cus04} using standard settings.  
\section{Spectral Classification and Spatial Distribution}

The classification was carried out using the guiding principles of near-infrared classification of WR stars according to \citet{cro06}, supplemented by the spectra that were taken at IRTF for stars of known type [WR152 WN3(h), WR127 (WN5o + O), WR138 (WN5+B), WR134 (WN6), WR120 (WN7), WR123 (WN8), WR108 (WN9h + OB); WR142 (WO2), WR143 (WC4 + OB?), WR111 (WC5), WR126 (WC5/WN), WR154 (WC6), WR137 (WC7pd + O), WR135 (WC8), WR121 (WC9d)]. These spectra of WR stars which have been well-studied in the optical- using the same set-up as the candidate WR stars - often helped decide borderline cases. The classification was made by eye, comparing nearby line pairs as in \citet{cro06}.  Ideally one should obtain EWs of the spectral lines, although in many cases this will be difficult, due to heavy blending for which the eye can readily compensate.
The numbers and subtypes of new WR stars found are:
WN5  2; WN6  6; WN7 5; WN8  3; WN9  1, for a total of 17 WN stars and
WC6 4 WC7 15; WC8 and WC8-9 22; WC9 13; for a total of 54 WC stars.
The grand total is 71 new WR stars, with 24\% WN and 76\% WC.

 It should be noted that the spectral differences among stars of type WC4-8 are subtle, and that uncertainties of one or even two subtypes are indicated by a colon in Tables 1-8.

\begin{deluxetable}{lccccccrrrrrrrr}
\label{tab:tab1}
\tabletypesize{\scriptsize}
\tablecaption{Previously Identified WR stars\label{tab:WRknown}}
\tablewidth{0pt}
\tablehead{
\colhead{Name} &
\colhead{$\alpha$(J2000)} &
\colhead{$\delta$(J2000)} &
\colhead{$l$ } &
\colhead{$b$}  &
\colhead{$B$\tablenotemark{a}}  &
\colhead{$V$\tablenotemark{a}} &
\colhead{$R$\tablenotemark{a}} &
\colhead{$J$\tablenotemark{a}} &
\colhead{$H$\tablenotemark{a}} &
\colhead{$K_{s}$\tablenotemark{a}}  \\
\colhead{(1)}  &
\colhead{(2)}  &
\colhead{(3)}  &
\colhead{(4)}  &
\colhead{(5)}  &
\colhead{(6)}  &
\colhead{(7)}  &
\colhead{(8)}  &
\colhead{(9)}  &
\colhead{(10)}  &
\colhead{(11)}  \\
}
\startdata
1059-62L    &    16 14 37.25    &    -51 26 26.4    &    331.81    &   -0.34    &    --    &    --    &    --    &    15.048    &    12.788   &    11.54    \\
1081-76L    &    16 24 58.87    &    -48 56 52.6    &    334.75    &   0.27    &    --    &    --    &    --    &    13.282    &    11.763   &    10.73    \\
1093-87L    &    16 31 29.21    &    -47 56 16.3    &    336.22    &   0.19    &    --    &    --    &    --    &    15.555    &    12.855   &    11.32    \\
1093-80L    &    16 31 49.06    &    -47 56 04.6    &    336.26    &   0.15    &    --    &    --    &    --    &    15.108    &    12.862   &    11.47    \\
1095-98L    &    16 35 23.21    &    -48 09 16.2     &    336.51    &   -0.43    &    --    &    --    &    --    &    14.900    &    12.770   &    11.43    \\
1218-38L    &    17 22 40.75    &    -35 04 52.8    &    352.20    &   0.74    &    --    &    --    &    --    &    15.105    &    12.044   &    10.33    \\
1385-9L    &    18 13 42.49    &    -17 28 12.3    &    13.15    &   0.13    &    --    &    --    &    20.02    &    11.205    &    9.699   &    8.57    \\
1425-15L    &    18 23 03.41    &    -13 10 00.5    &    18.01    &   0.18    &    15.23    &    14.49    &    14.41    &    10.339    &   9.282    &    8.27    \\
1428-157L    &    18 25 53.09    &    -13 28 32.5    &    18.05    &   -0.57    &    16.07    &    15.13    &    --    &    10.317    &   9.521    &    8.96    \\
1505-86L    &    18 41 48.47    &    -04 00 12.8    &    28.27    &   0.31    &    --    &    --    &    --    &    15.618    &    13.323   &    11.99    \\
1613-50L    &    19 06 36.53    &    +07 29 52.4    &    41.33    &   0.06    &    --    &    --    &    --    &    14.237    &    12.65   &    11.61    \\
1671-32L    &    19 20 40.40    &    +13 50 35.1    &    48.55    &   -0.05    &    --    &    --    &    --    &    13.573    &    11.804   &    10.76    \\\hline
\enddata
\tablecomments{Previously identified Wolf-Rayet stars from \citet{sha09} that were observed using SpeX.  A lack of BVR data implies that the star is below the 21st magnitude plate limits of the digitized sky surveys. }
\tablenotetext{a}{The $B$,$V$,and $R$ photometry comes from the NOMAD catalog while $J$, $H$, and $K_{s}$ photometry comes from 2MASS}

\end{deluxetable}

\begin{deluxetable}{lccccccrrrrrrrr}
\label{tab:tab1}
\tabletypesize{\scriptsize}
\tablecaption{New WR stars\label{tab:WRnew}}
\tablewidth{0pt}
\tablehead{
\colhead{Name} &
\colhead{$\alpha$(J2000)} &
\colhead{$\delta$(J2000)} &
\colhead{$l$ } &
\colhead{$b$}  &
\colhead{$B$\tablenotemark{a}}  &
\colhead{$V$\tablenotemark{a}} &
\colhead{$R$\tablenotemark{a}} &
\colhead{$J$\tablenotemark{a}} &
\colhead{$H$\tablenotemark{a}} &
\colhead{$K_{s}$\tablenotemark{a}}  \\
\colhead{(1)}  &
\colhead{(2)}  &
\colhead{(3)}  &
\colhead{(4)}  &
\colhead{(5)}  &
\colhead{(6)}  &
\colhead{(7)}  &
\colhead{(8)}  &
\colhead{(9)}  &
\colhead{(10)}  &
\colhead{(11)}  \\
}
\startdata
1023-63L    &    15 52 09.48    &    -54 17 14.5     &    327.39    &   -0.23    &    --    &    --    &    --    &    16.13    &    15.06   &    14.37    \\
1042-25L    &    16 00 25.25    &    -52 03 29.6    &    329.77    &   0.68    &    --    &    --    &    --    &    12.02    &    10.65   &    9.88    \\
1038-22L    &    16 00 26.41    &    -52 11 10.1    &    329.69    &   0.58    &    --    &    --    &    19.70    &    11.53    &    10.20   &    9.29    \\
1054-43L    &    16 10 06.26    &    -50 47 58.6    &    331.74    &   0.61    &    --    &    --    &    --    &    15.62    &    13.05   &    11.53    \\
1051-67L    &    16 10 06.67    &    -51 47 24.5    &    331.07    &   -0.11    &    --    &    --    &    --    &    14.86    &    12.63   &    11.24    \\
1077-55L    &    16 24 22.70    &    -49 00 42.3    &    334.63    &   0.30    &    --    &    --    &    --    &    15.15    &    13.13   &    11.97    \\
1085-72L    &    16 27 42.39    &    -48 30 34.2    &    335.37    &   0.25    &    --    &    --    &    --    &    14.50    &    12.37   &    11.21    \\
1085-69L    &    16 28 40.25    &    -48 18 12.9    &    335.63    &   0.28    &    --    &    --    &    --    &    14.73    &    13.16   &    11.50    \\
1085-83L    &    16 29 35.82    &    -48 19 34.2    &    335.72    &   0.15    &    --    &    --    &    --    &    16.70    &    13.49   &    11.83    \\
1093-138L    &    16 32 15.22    &    -47 56 12.7    &    336.31    &   0.10    &    --    &    --    &    --    &    16.00    &    14.17   &    12.75    \\
1093-140LB    &    16 32 47.99        &    -47 44 53.2    &    336.51   &    0.16    &    --    &    17.74    &    --    &    15.79    &   14.29    &    13.99    \\
1093-140L    &    16 32 49.78    &    -47 44 31.4    &    336.52    &   0.16    &    --    &    --    &    --    &    16.15    &    13.97   &    12.28    \\
1091-46L    &    16 33 14.06    &    -48 17 37.2    &    336.16    &   -0.26    &    --    &    --    &    --    &    13.93    &    11.76   &    10.02    \\
1093-59L    &    16 33 45.45    &    -47 51 29.1    &    336.54    &   -0.03    &    --    &    --    &    --    &    15.56    &    12.98   &    11.41    \\
1095-189L    &    16 33 48.13    &    -47 52 52.8    &    336.53    &   -0.05    &    --    &    --    &    --    &    10.43    &    9.62   &    9.35    \\
1097-156L    &    16 34 57.45    &    -47 04 13.0    &    337.26    &   0.35    &    --    &    --    &    --    &    13.75    &    11.71   &    10.46    \\
1097-71L    &    16 35 44.37    &    -47 19 42.2    &    337.16    &   0.08    &    --    &    --    &    --    &    15.28    &    13.72   &    12.04    \\
1097-34L    &    16 35 51.16    &    -47 19 51.3    &    337.17    &   0.06    &    --    &    --    &    --    &    13.60    &    11.67   &    10.39    \\
1106-31L    &    16 37 24.00    &    -46 26 28.6    &    338.00    &   0.47    &    17.78    &    15.29    &    13.31    &    10.27    &   9.59    &    8.93    \\
1105-76L    &    16 38 20.18    &    -46 23 43.8    &    338.14    &   0.38    &    --    &    --    &    --    &    14.77    &    12.74   &    11.48    \\
1109-74L    &    16 40 17.12    &    -46 20 09.7    &    338.41    &   0.17    &    --    &    --    &    --    &    16.92    &    13.10   &    11.25    \\
1115-197L    &    16 43 40.36    &    -45 57 57.5    &    339.08    &   -0.03    &    18.20    &    16.73    &    14.59    &    10.56    &   9.71    &    9.14    \\
1138-133L    &    16 51 19.32    &    -43 26 55.3    &    341.88    &   0.56    &    --    &    --    &    --    &    13.56    &    11.93   &    10.95    \\
1133-59L    &    16 51 29.69    &    -43 53 35.3    &    341.55    &   0.25    &    --    &    --    &    --    &    14.73    &    13.46   &    12.06    \\
1168-91L    &    17 09 32.64    &    -41 29 47.3    &    345.48    &   -0.88    &    --    &    --    &    --    &    15.13    &    13.43   &    11.84    \\
1179-129L    &    17 11 00.84    &    -39 49 31.2    &    346.99    &   -0.12    &    --    &    --    &    --    &    15.26    &    13.89   &    12.81    \\
1181-82L    &    17 11 28.52    &    -39 13 16.8    &    347.53    &   0.17    &    --    &    --    &    --    &    13.98    &    12.20   &    10.98    \\
1181-81L    &    17 11 36.12    &    -39 11 07.9    &    347.58    &   0.17    &    --    &    --    &    --    &    13.35    &    12.08   &    10.81    \\
1181-211L    &    17 11 46.14    &    -39 20 27.7    &    347.47    &   0.05    &    20.75    &    17.70    &    16.33    &    10.88    &   10.00    &    9.49    \\
1189-110L    &    17 14 09.56    &    -38 11 20.9    &    348.68    &   0.35    &    --    &    --    &    --    &    14.34    &    12.78   &    11.59    \\
1245-23L    &    17 33 33.22    &    -32 36 16.4    &    355.52    &   0.24    &    --    &    --    &    --    &    15.99    &    12.59   &    10.76    \\
1269-166L    &    17 41 13.51    &    -30 03 41.1    &    358.54    &   0.22    &    --    &    --    &    --    &    13.52    &    11.70   &    10.56    \\
1275-184L    &    17 44 06.89    &    -30 01 13.2    &    358.90    &   -0.29    &    --    &    --    &    --    &    13.90    &    11.52   &    10.17    \\
1322-220L    &    17 55 20.21    &    -24 07 38.2    &    5.24    &   0.60    &    19.93    &    --    &    16.80    &    11.84    &   10.95    &    10.32    \\
1327-25L    &    17 59 02.86    &    -24 20 51.12    &    5.48    &   -0.24    &    --    &    --    &    --    &    13.78    &    12.38   &    10.89    \\
1342-208L    &    17 59 48.22    &    -22 14 52.4    &    7.38    &   0.65    &    --    &    --    &    17.71    &    11.40    &    10.29   &    9.47    \\
1381-20L    &    18 12 57.27    &    -18 01 20.7    &    12.58    &   0.02    &    --    &    --    &    --    &    14.24    &     13.35   &    10.75    \\
1395-86L    &    18 16 02.36    &    -16 53 59.4    &    13.92    &   -0.09    &    --    &    --    &    --    &    18.08    &    13.96   &    11.85    \\
1434-43L    &    18 23 32.32    &    -12 03 58.5    &    19.03    &   0.59    &    --    &    --    &    --    &    14.45    &    12.90   &    11.69    \\
1431-34L    &    18 25 53.63    &    -12 50 03.0    &    18.62    &   -0.27    &    20.65    &    --    &    17.43    &    11.53    &   10.13    &    9.28    \\
1463-7L    &    18 33 47.64    &    -09 23 07.7    &    22.58    &   -0.39    &    --    &    --    &    --    &    12.18    &    10.52   &    9.36    \\
1477-55L    &    18 35 47.67    &    -07 17 50.1    &    24.66    &   0.13    &    --    &    --    &    --    &    15.91    &    12.89   &    11.01    \\
1487-80L    &    18 38 00.49    &    -06 26 46.1    &    25.67    &   0.03    &    --    &    --    &    --    &    15.63    &    13.02   &    11.29    \\
1483-212L    &    18 38 27.16    &    -07 10 45.0    &    25.07    &   -0.40    &    --    &    --    &    --    &    13.64    &    11.73   &    10.59    \\
1489-36L    &    18 38 38.94    &    -06 00 16.0    &    26.13    &   0.10    &    --    &    --    &    --    &    14.78    &    13.04   &    11.15    \\
1493-9L    &    18 39 34.58    &    -05 44 23.2    &    26.47    &   0.01    &    18.36    &    --    &    16.96    &    11.83    &   10.49    &    9.56    \\
1487-212L    &    18 39 42.53    &    -06 41 46.4    &    25.64    &   -0.46    &    --    &    --    &    --    &    13.27    &    11.50   &    10.50    \\
1495-32L    &    18 41 23.36    &    -05 40 58.1    &    26.73    &   -0.36    &    17.97    &    17.39    &    16.16    &    12.35    &   11.15    &    10.25    \\
1503-160L    &    18 41 34.06    &    -05 04 01.4    &    27.30    &   -0.12    &    19.18    &    --    &    15.71    &    10.22    &   9.21    &    8.51    \\
1513-111L    &    18 43 17.27    &    -03 20 23.6    &    29.03    &   0.29    &    --    &    --    &    --    &    16.17    &    14.09   &    12.04    \\
1522-55L    &    18 43 39.65    &    -02 29 35.9    &    29.83    &   0.59    &    --    &    --    &    --    &    13.41    &    12.24   &    11.47    \\
1517-138L    &    18 43 58.03    &    -02 45 17.1    &    29.63    &   0.40    &    17.96    &    16.25    &    14.30    &    10.00    &   9.16    &    8.53    \\
1527-13L    &    18 47 38.33    &    -02 06 38.9    &    30.62    &   -0.12    &    --    &    --    &    --    &    16.33    &    12.70   &    10.56    \\
1528-15L    &    18 49 32.31    &    -02 24 27.0    &    30.57    &   -0.68    &    --    &    --    &    --    &    14.13    &    12.13   &    10.66    \\
1536-180L    &    18 51 10.77    &    -01 30 03.4    &    31.57    &   -0.63    &    17.26    &    15.51    &    14.58    &    10.42    &   9.76    &    9.34    \\
1551-19L    &    18 52 32.97    &    +00 14 26.8    &    33.27    &   -0.14    &    --    &    --    &    --    &    16.91    &    13.62   &    11.78    \\
1563-66L    &    18 55 44.44    &    +01 36 43.9    &    34.86    &   -0.22    &    --    &    --    &    --    &    16.56    &    13.29   &    11.45    \\
1563-89L    &    18 56 02.04    &    +01 36 32.9    &    34.89    &   -0.29    &    --    &    --    &    --    &    17.13    &    14.32   &    12.55    \\
1567-51L    &    18 56 07.90    &    +02 20 49.0    &    35.56    &   0.03    &    --    &    --    &    --    &    14.72    &    12.27   &    10.87    \\
1583-64L    &    19 00 59.99    &    +03 55 35.6    &    37.52    &   -0.33    &    --    &    --    &    --    &    17.49    &    14.80   &    12.79    \\
1583-48L    &    19 01 26.62    &    +03 51 55.5    &    37.51    &   -0.46    &    --    &    --    &    --    &    14.70    &    12.67   &    11.25    \\
1583-47L    &    19 01 27.11    &    +03 51 54.4    &    37.51    &   -0.46    &    --    &    --    &    --    &    14.25    &    12.27   &    10.99    \\
1603-75L    &    19 04 33.49    &    +06 05 18.5    &    39.84    &   -0.13    &    --    &    --    &    --    &    16.31    &    14.38   &    13.68    \\
1650-96L    &    19 13 24.01    &    +11 43 24.3    &    45.85    &   0.53    &    --    &    --    &    --    &    8.61    &    7.99    &   7.80    \\
1657-51L    &    19 16 18.38    &    +12 46 49.2    &    47.12    &   0.39    &    --    &    --    &    --    &    12.96    &    11.82   &    10.77    \\
1670-57L    &    19 17 32.79    &    +14 08 27.9    &    48.46    &   0.76    &    --    &    --    &    19.93    &    13.90    &    12.72   &    11.67    \\
1652-24L    &    19 17 41.21    &    +11 29 18.9    &    46.13    &   -0.51    &    --    &    --    &    --    &    15.29    &    13.00   &    11.53    \\
1669-24L    &    19 18 31.71    &    +13 43 17.9    &    48.20    &   0.36    &    --    &    --    &    --    &    14.35    &    12.59   &    11.33    \\
1675-17L    &    19 22 53.61    &    +14 08 50.0    &    49.07    &   -0.38    &    --    &    --    &    --    &    12.67    &    10.91   &    9.68    \\
1675-10L    &    19 22 54.45    &    +14 11 27.9    &    49.11    &   -0.36    &    --    &    --    &    --    &    12.83    &    11.02   &    9.58    \\
1698-70L    &    19 24 46.90    &    +17 14 25.0    &    52.01    &   0.68    &    --    &    --    &    --    &    12.65    &    11.19   &    10.25    \\
\hline
\enddata
\tablenotetext{a}{The $B$,$V$, and $R$ photometry comes from the NOMAD catalog while $J$, $H$, and $K_{s}$ photometry comes from 2MASS}
\end{deluxetable}

\begin{deluxetable}{lccccccrrrrrrrr}
\label{tab:tab1}
\tabletypesize{\scriptsize}
\tablecaption{Known WR stars\label{tab:WRnew}}
\tablewidth{0pt}
\tablehead{
\colhead{Name} &
\colhead{Subtype\tablenotemark{a}} &
\colhead{A$\frac{J-K_{s}}{K_{s}}$\tablenotemark{b}} &
\colhead{A$\frac{H-K_{s}}{K_{s}}$\tablenotemark{b}} &
\colhead{A$_{K_{s}}$}  &
\colhead{K$_{s}$}  &
\colhead{M$_{K_{s}}$\tablenotemark{c}} &
\colhead{DM} &
\colhead{$d$\tablenotemark{d}} &
\colhead{R$_{G}$\tablenotemark{d}}  \\
\colhead{(1)}  &
\colhead{(2)}  &
\colhead{(3)}  &
\colhead{(4)}  &
\colhead{(5)}  &
\colhead{(6)}  &
\colhead{(7)}  &
\colhead{(8)}  &
\colhead{(9)}  &
\colhead{(10)}  \\
}
\startdata
1059-62L    &    WC8:    &    2.06    &    1.59    &    1.82    &   11.54    &    -4.65    &    14.36    &    7.46    &    4.01    \\
1081-76L    &    WC8    &    1.42    &    1.19    &    1.31    &   10.73    &    -4.65    &    14.07    &    6.53    &    3.81    \\
1093-87L    &    WC8:    &    2.55    &    2.10    &    2.33    &   11.32    &    -4.65    &    13.64    &    5.36    &    4.20    \\
1093-80L    &    WC8    &    2.15    &    1.85    &    2.00    &   11.47    &    -4.65    &    14.12    &    6.66    &    3.60    \\
1095-98L    &    WC8    &    2.04    &    1.74    &    1.89    &   11.43    &    -4.65    &    14.19    &    6.89    &    3.51    \\
1218-38L    &    WC8    &    2.91    &    2.43    &    2.67    &   10.33    &    -4.65    &    12.31    &    2.89    &    5.65    \\
1385-9L    &    WC8    &    1.48    &    1.37    &    1.42    &   8.57    &    -4.65    &    11.80    &    2.29    &    6.29    \\
1425-15L    &    WC6:    &    0.97    &    0.79    &    0.88    &   8.27    &    -4.59    &    11.97    &    2.48    &    6.19    \\
1428-157L    &    WN6    &    0.66    &    0.53    &    0.59    &   8.96    &    -4.41    &    12.78    &    3.59    &    5.21    \\
1505-86L    &    WC7:    &    2.01    &    1.37    &    1.69    &   11.99    &    -4.59    &    14.89    &    9.52    &    4.51    \\
1613-50L    &    WC4::    &    ---    &    ---    &    ---    &   11.61    &    ---    &    ---    &    ---    &    ---    \\
1671-32L    &    WC7:    &    1.47    &    0.84    &    1.15    &   10.76    &    -4.59    &    14.20    &    6.91    &    6.50    \\\enddata
\tablenotetext{a}{Differences among stars of type WC4-8 are difficult to distinguish from one another.  A colon (:) indicates an uncertainty of up to $\pm$2 subtypes.}
\tablenotetext{b}{Extinction was calculated from 2MASS colors and subtype values provided in \citet{cro06}}
\tablenotetext{c}{M$_{K_{s}}$ values are derived for spectral subtypes by \citet{cro06}}
\tablenotetext{d}{Distances (d) and Galactocentric radius (R$_{G}$) reported in kpc with typical uncertainties of 25\%}

\end{deluxetable}

\begin{deluxetable}{lccccccrrrrrrrr}
\label{tab:tab1}
\tabletypesize{\scriptsize}
\tablecaption{New WR stars\label{tab:WRnew}}
\tablewidth{0pt}
\tablehead{
\colhead{Name} &
\colhead{Subtype\tablenotemark{a}} &
\colhead{A$\frac{J-K_{s}}{K_{s}}$\tablenotemark{b}} &
\colhead{A$\frac{H-K_{s}}{K_{s}}$\tablenotemark{b}} &
\colhead{A$_{K_{s}}$}  &
\colhead{K$_{s}$}  &
\colhead{M$_{K_{s}}$\tablenotemark{c}} &
\colhead{DM} &
\colhead{$d$\tablenotemark{d}} &
\colhead{R$_{G}$\tablenotemark{d}}  \\
\colhead{(1)}  &
\colhead{(2)}  &
\colhead{(3)}  &
\colhead{(4)}  &
\colhead{(5)}  &
\colhead{(6)}  &
\colhead{(7)}  &
\colhead{(8)}  &
\colhead{(9)}  &
\colhead{(10)}  \\
}
\startdata

1023-63L    &    WC7:    &    0.76    &    0.19    &    0.48    &   14.37    &    -4.59    &    18.49    &    49.80    &    42.88    \\
1042-25L    &    WN8    &    1.34    &    1.20    &    1.27    &   9.88    &    -5.92    &    14.53    &    8.05    &    4.34    \\
1038-22L    &    WC7:    &    1.08    &    0.60    &    0.84    &   9.29    &    -4.59    &    13.04    &    4.05    &    5.40    \\
1054-43L    &    WC9    &    2.59    &    2.30    &    2.45    &   11.53    &    ---    &    ---    &    ---    &    ---    \\
1051-67L    &    WC7:    &    2.01    &    1.46    &    1.73    &   11.24    &    -4.59    &    14.10    &    6.61    &    4.20    \\
1077-55L    &    WC6:    &    1.71    &    1.06    &    1.39    &   11.97    &    -4.59    &    15.18    &    10.84    &    4.82    \\
1085-72L    &    WC8-9    &    1.92    &    1.41    &    1.66    &   11.21    &    -4.65    &    14.20    &    6.92    &    3.63    \\
1085-69L    &    WC8    &    1.88    &    2.34    &    2.11    &   11.50    &    -4.65    &    14.04    &    6.44    &    3.74    \\
1085-83L    &    WC8:    &    2.98    &    2.34    &    2.66    &   11.83    &    -4.65    &    13.82    &    5.80    &    4.00    \\
1093-138L    &    WC8:    &    1.89    &    1.89    &    1.89    &   12.75    &    -4.65    &    15.51    &    12.65    &    5.94    \\
1093-140LB    &    WN9    &    1.12    &    0.35    &    0.73    &   13.99    &    -5.92    &    19.18    &    68.40    &    60.70    \\
1093-140L    &    WC7:    &    2.18    &    2.02    &    2.10    &   12.28    &    -4.59    &    14.76    &    8.97    &    3.58    \\
1091-46L    &    WC8    &    2.33    &    2.48    &    2.41    &   10.02    &    -4.65    &    12.26    &    2.83    &    6.02    \\
1093-59L    &    WC9+late-type spectrum    &    2.62    &    2.39    &   2.51    &    11.41    &    ---    &    ---    &    ---    &    ---    \\
1095-189L    &    WC7:    &    0.31    &    -0.56    &    -0.13    &   9.35    &    -4.59    &    14.06    &    6.50    &    3.63    \\
1097-156L    &    WN6:    &    2.08    &    1.97    &    2.03    &   10.46    &    -4.41    &    12.84    &    3.71    &    5.28    \\
1097-71L    &    WC9    &    2.02    &    2.59    &    2.31    &   12.04    &    ---    &    ---    &    ---    &    ---    \\
1097-34L    &    WC8    &    1.86    &    1.64    &    1.75    &   10.39    &    -4.65    &    13.28    &    4.54    &    4.66    \\
1106-31L    &    WC9    &    0.74    &    0.73    &    0.74    &   8.93    &    ---    &    ---    &    ---    &    ---    \\
1105-76L    &    WC8    &    1.91    &    1.59    &    1.75    &   11.48    &    -4.65    &    14.38    &    7.51    &    3.19    \\
1109-74L    &    WC7:    &    3.38    &    2.32    &    2.85    &   11.25    &    -4.59    &    12.99    &    3.96    &    5.03    \\
1115-197L    &    WN6    &    0.83    &    0.74    &    0.79    &   9.14    &    -4.41    &    12.76    &    3.57    &    5.32    \\
1138-133L    &    WN6    &    1.63    &    1.51    &    1.57    &   10.95    &    -4.41    &    13.79    &    5.72    &    3.55    \\
1133-59L    &    WC9    &    1.64    &    2.07    &    1.85    &   12.06    &    ---    &    ---    &    ---    &    ---    \\
1168-91L    &    WC7:    &    1.79    &    1.83    &    1.81    &   11.84    &    -4.59    &    14.62    &    8.40    &    2.14    \\
1179-129L    &    WC6:    &    1.22    &    0.91    &    1.07    &   12.81    &    -4.59    &    16.34    &    18.52    &    10.41    \\
1181-82L    &    WC8    &    1.72    &    1.53    &    1.63    &   10.98    &    -4.65    &    14.00    &    6.32    &    2.70    \\
1181-81L    &    WC8    &    1.41    &    1.62    &    1.52    &   10.81    &    -4.65    &    13.94    &    6.15    &    2.82    \\
1181-211L    &    WN7:    &    0.68    &    0.43    &    0.55    &   9.49    &    -4.77    &    13.71    &    5.51    &    3.34    \\
1189-110L    &    WC9    &    1.69    &    1.69    &    1.69    &   11.59    &    ---    &    ---    &    ---    &    ---    \\
1245-23L    &    WC9    &    3.35    &    2.86    &    3.10    &   10.76    &    ---    &    ---    &    ---    &    ---    \\
1269-166L    &    WC8    &    1.70    &    1.39    &    1.54    &   10.56    &    -4.65    &    13.67    &    5.41    &    3.09    \\
1275-184L    &    WN8    &    2.41    &    2.25    &    2.33    &   10.17    &    -5.92    &    13.76    &    5.66    &    2.85    \\
1322-220L    &    WN5    &    0.90    &    0.86    &    0.88    &   10.32    &    -4.41    &    13.85    &    5.88    &    2.70\\
1327-25L    &    WC9    &    1.78    &    2.24    &    2.01    &   10.89    &    ---    &    ---    &    ---    &    ---    \\
1342-208L    &    WN6    &    1.17    &    1.19    &    1.18    &   9.47    &    -4.41    &    12.70    &    3.47    &    5.07    \\
1381-20L    &    WC9    &    2.18    &    4.26    &    3.22    &   10.75    &    ---    &    ---    &    ---    &    ---    \\
1395-86L    &    WC8    &    3.89    &    3.15    &    3.52    &   11.85    &    -4.65    &    12.97    &    3.94    &    4.77    \\
1434-43L    &    WC8    &    1.56    &    1.51    &    1.53    &   11.69    &    -4.65    &    14.81    &    9.15    &    2.99    \\
1431-34L    &    WN8    &    1.42    &    1.34    &    1.38    &   9.28    &    -5.92    &    13.82    &    5.82    &    3.52    \\
1463-7L    &    WC8    &    1.60    &    1.43    &    1.51    &   9.36    &    -4.65    &    12.50    &    3.16    &    5.72    \\
1477-55L    &    WC9    &    3.13    &    2.94    &    3.03    &   11.01    &    ---    &    ---    &    ---    &    ---    \\
1487-80L    &    WC9    &    2.75    &    2.68    &    2.71    &   11.29    &    ---    &    ---    &    ---    &    ---    \\
1483-212L    &    WN7:    &    1.80    &    1.58    &    1.69    &   10.59    &    -4.77    &    13.67    &    5.42    &    4.26    \\
1489-36L    &    WC9    &    2.28    &    2.97    &    2.62    &   11.15    &    ---    &    ---    &    ---    &    ---    \\
1493-9L    &    WC8    &    1.23    &    1.00    &    1.12    &   9.56    &    -4.65    &    13.09    &    4.14    &    5.14    \\
1487-212L    &    WN7    &    1.61    &    1.32    &    1.46    &   10.50    &    -4.77    &    13.81    &    5.78    &    4.13    \\
1495-32L    &    WC8    &    1.12    &    0.94    &    1.03    &   10.25    &    -4.65    &    13.87    &    5.94    &    4.16    \\
1503-160L    &    WN7    &    0.90    &    0.78    &    0.84    &   8.51    &    -4.77    &    12.44    &    3.08    &    5.93    \\
1513-111L    &    WC7:    &    2.35    &    2.68    &    2.52    &   12.04    &    -4.59    &    14.11    &    6.63    &    4.20    \\
1522-55L    &    WC9    &    1.15    &    0.92    &    1.03    &   11.47    &    ---    &    ---    &    ---    &    ---    \\
1517-138L    &    WN7    &    0.74    &    0.65    &    0.69    &   8.53    &    -4.77    &    12.61    &    3.33    &    5.84    \\
1527-13L    &    WC8    &    3.57    &    3.19    &    3.38    &   10.56    &    -4.65    &    11.83    &    2.33    &    6.61    \\
1528-15L    &    WC7    &    1.91    &    1.62    &    1.76    &   10.66    &    -4.59    &    13.49    &    4.99    &    4.91    \\
1536-180L    &    WN5    &    0.60    &    0.46    &    0.53    &   9.34    &    -4.41    &    13.22    &    4.40    &    5.28    \\
1551-19L    &    WC8:    &    3.15    &    2.66    &    2.90    &   11.78    &    -4.65    &    13.53    &    5.08    &    5.09    \\
1563-66L    &    WC8:    &    3.13    &    2.65    &    2.89    &   11.45    &    -4.65    &    13.21    &    4.39    &    5.50    \\
1563-89L    &    WC7:    &    2.66    &    2.16    &    2.41    &   12.55    &    -4.59    &    14.73    &    8.83    &    5.20    \\
1567-51L    &    WC7:    &    2.17    &    1.49    &    1.83    &   10.87    &    -4.59    &    13.63    &    5.33    &    5.19    \\
1583-64L    &    WC7:    &    2.74    &    2.62    &    2.68    &   12.79    &    -4.59    &    14.70    &    8.70    &    5.53    \\
1583-48L    &    WC8    &    2.02    &    1.89    &    1.96    &   11.25    &    -4.65    &    13.95    &    6.16    &    5.21    \\
1583-47L    &    WC8    &    1.89    &    1.63    &    1.76    &   10.99    &    -4.65    &    13.88    &    5.96    &    5.24    \\
1603-75L    &    WC8    &    1.47    &    0.58    &    1.02    &   13.68    &    -4.65    &    17.31    &    28.97    &    23.09    \\
1650-96L    &    WN6    &    0.42    &    0.05    &    0.23    &   7.80    &    -4.41    &    11.98    &    2.49    &    7.00    \\
1657-51L    &    WC7    &    1.05    &    0.86    &    0.95    &   10.77    &    -4.59    &    14.40    &    7.60    &    6.49    \\
1670-57L    &    WC6:    &    1.08    &    0.87    &    0.98    &   11.67    &    -4.59    &    15.28    &    11.38    &    8.57    \\
1652-24L    &    WC7:    &    2.10    &    1.61    &    1.86    &   11.53    &    -4.59    &    14.27    &    7.14    &    6.25    \\
1669-24L    &    WC6:    &    1.61    &    1.24    &    1.43    &   11.33    &    -4.59    &    14.49    &    7.91    &    6.72    \\
1675-17L    &    WC7:    &    1.59    &    1.20    &    1.39    &   9.68    &    -4.59    &    12.87    &    3.75    &    6.67    \\
1675-10L    &    WC8    &    1.89    &    1.93    &    1.91    &   9.58    &    -4.65    &    12.33    &    2.92    &    6.95    \\
1698-70L    &    WN6    &    1.36    &    1.21    &    1.29    &   10.25    &    -4.41    &    13.37    &    4.73    &    6.72    \\
\enddata
\tablenotetext{a}{Differences among stars of type WC4-8 are difficult to distinguish from one another.  A colon (:) indicates an uncertainty of $\pm$2 subtypes.}
\tablenotetext{b}{Extinction was calculated from 2MASS colors and subtype values provided in \citet{cro06}}
\tablenotetext{c}{M$_{K_{s}}$ values are derived for spectral subtypes by \citet{cro06}}
\tablenotetext{d}{Distances (d) and Galactocentric radius (R$_{G}$) reported in kpc with typical uncertainties of 25\%}

\end{deluxetable}

\begin{deluxetable}{lccccccrrrrrrrr}
\label{tab:tab1}
\tabletypesize{\scriptsize}
\tablecaption{Equivalent Width ($\AA$) Measurement for the Most Prominent Lines of the New WN Stars\label{tab:WNnew}}
\tablewidth{0pt}
\tablehead{
\colhead{Name} &
\colhead{SpT}&
\colhead{N V} &
\colhead{He I}  &
\colhead{He II + Br $\gamma$}  &
\colhead{He II} &
\colhead{$\frac{W_{2.189}}{W_{2.165}}$} \\
&
&
\colhead{2.100 $\micron$m}&
\colhead{2.115 $\micron$m}&
\colhead{2.165 $\micron$m}&
\colhead{2.189 $\micron$m}\\
&
&
\colhead{$\AA$}&
\colhead{$\AA$}&
\colhead{$\AA$}&
\colhead{$\AA$}\\
\colhead{(1)}  &
\colhead{(2)}  &
\colhead{(3)}  &
\colhead{(4)}  &
\colhead{(5)}  &
\colhead{(6)}  &
\colhead{(7)}   \\
}
\startdata
1322-220L    &    WN5    &    -8    &    -10    &    -32    &    -65       &    2.0    \\
1536-180L    &    WN5    &    -3    &    -10    &    -17    &    -55       &    3.2    \\
1097-156L    &    WN6:    &    ---    &    -62    &    -67    &   -108        &    1.6    \\
1115-197L    &    WN6    &    -3    &    -27    &    -38    &    -68       &    1.8    \\
1138-133L    &    WN6    &    -2    &    -29    &    -9    &    -104       &    11.6    \\
1342-208L    &    WN6    &    -4    &    -41    &    -13    &    -72       &    5.5    \\
1650-96L    &    WN6    &    -5    &    -37    &    -49    &    -52       &    1.1    \\
1698-70L    &    WN6    &    ---    &    -24    &    -9    &    -101       &    11.2    \\
1181-211L    &    WN7:    &    ---    &    -16    &    -21    &    -22       &    1.0    \\
1483-212L    &    WN7:    &    -6    &    -62    &    -60    &    -51       &    0.9    \\
1487-212L    &    WN7    &    -7    &    -60    &    -63    &    -48       &    0.8    \\
1503-160L    &    WN7    &    ---    &    -53    &    -52    &    -32       &    0.6    \\
1517-138L    &    WN7    &    -7    &    -56    &    -61    &    -33       &    0.5    \\
1042-25L    &    WN8    &    -3    &    -41    &    -63    &    -17       &    0.3    \\
1275-184L    &    WN8    &    -4    &    -51    &    -59    &    -14       &    0.2    \\
1431-34L    &    WN8    &    -5    &    -44    &    -50    &    -20       &    0.4    \\
1093-140LB    &    WN9    &    ---    &    -101    &    -90    &   ---        &    ---    \\\enddata
\end{deluxetable}

\begin{deluxetable}{lccccccrrrrrrrr}
\label{tab:tab1}
\tabletypesize{\scriptsize}
\tablecaption{Equivalent Width ($\AA$) Measurement for the Most Prominent Lines of the New WC Stars\label{tab:WCnew}}
\tablewidth{0pt}
\tablehead{
\colhead{Name} &
\colhead{SpT}&
\colhead{C IV} &
\colhead{He I + C III}  &
\colhead{He I}  &
\colhead{He II} &
\colhead{$\frac{W_{2.076}}{W_{2.110}}$} \\
&
&
\colhead{2.076 $\micron$m}&
\colhead{2.110 $\micron$m}&
\colhead{2.165 $\micron$m}&
\colhead{2.189 $\micron$m}\\
&
&
\colhead{$\AA$}&
\colhead{$\AA$}&
\colhead{$\AA$}&
\colhead{$\AA$}\\
\colhead{(1)}  &
\colhead{(2)}  &
\colhead{(3)}  &
\colhead{(4)}  &
\colhead{(5)}  &
\colhead{(6)}  &
\colhead{(7)}   \\
}
\startdata
1077-55L    &    WC6:    &    -584    &    -101    &    -16    &   -65    &    5.8    \\
1179-129L    &    WC6:    &    -957    &    -167    &    ---    &   ---    &    5.7    \\
1670-57L    &    WC6:    &    -969    &    -128    &    -25    &   -89    &    7.6    \\
1669-24L    &    WC6:    &    -190, -423    &    -25, -47    &   -90,-25    &    -6,-56    &    ---    \\
1023-63L    &    WC7:    &    -794    &    -165    &    -13    &   -41    &    4.8    \\
1038-22L    &    WC7:    &    -278    &    -69    &    -40    &    -51   &    4.0    \\
1051-67L    &    WC7:    &    -465    &    -111    &    ---    &   -20    &    4.2    \\
1093-140L    &    WC7:    &    -372    &    -62    &    ---    &   ---    &    6.0    \\
1095-189L    &    WC7:    &    -792    &    -180    &    ---    &   ---    &    4.4    \\
1109-74L    &    WC7:    &    -300    &    -33    &    -5    &    -55   &    9.1    \\
1168-91L    &    WC7:    &    -544    &    -88    &    ---    &    -55   &    6.2    \\
1513-111L    &    WC7:    &    -494    &    -89    &    -7    &    -58   &    5.6    \\
1528-15L    &    WC7    &    -827    &    -179    &    ---    &    -44   &    4.6    \\
1563-89L    &    WC7:    &    -756    &    -165    &    -10    &   -49    &    4.6    \\
1567-51L    &    WC7:    &    -272    &    -40    &    ---    &    -21   &    6.8    \\
1583-64L    &    WC7:    &    -779    &    -160    &    ---    &   -82    &    4.9    \\
1657-51L    &    WC7    &    -530    &    -86    &    -13    &    -48   &    6.2    \\
1652-24L    &    WC7:    &    -690    &    -88    &    -20    &    -68   &    7.8    \\
1675-17L    &    WC7:    &    -680    &    -111    &    -33    &   -66    &    6.1    \\
1085-69L    &    WC8    &    -146    &    -111    &    -17    &    -26   &    1.3    \\
1085-83L    &    WC8:    &    -321    &    -100    &    -28    &   -32    &    3.2    \\
1093-138L    &    WC8:    &    -557    &    ---    &    ---    &   ---    &    ---    \\
1091-46L    &    WC8    &    -239    &    -134    &    -10    &    -18   &    1.8    \\
1097-34L    &    WC8    &    -202    &    -141    &    -19    &    -32   &    1.4    \\
1105-76L    &    WC8    &    -380    &    -116    &    -31    &    -55   &    3.3    \\
1181-82L    &    WC8    &    -258    &    -121    &    ---    &    -27   &    2.1    \\
1181-81L    &    WC8    &    -287    &    -100    &    -20    &    -35   &    2.9    \\
1269-166L    &    WC8    &    -197    &    -113    &    -59    &   -45    &    1.7    \\
1395-86L    &    WC8    &    -136    &    -91    &    -18    &    -25   &    1.5    \\
1434-43L    &    WC8    &    -461    &    -161    &    -26    &    -43   &    2.9    \\
1463-7L    &    WC8    &    -250    &    -158    &    -69    &    -51   &    1.6    \\
1493-9L    &    WC8    &    -254    &    -104    &    -9    &    -26   &    2.4    \\
1495-32L    &    WC8    &    -266    &    -107    &    -46    &    -47   &    2.5    \\
1527-13L    &    WC8    &    -176    &    -48    &    -9    &    -62   &    3.7    \\
1551-19L    &    WC8:    &    -374    &    -108    &    ---    &   -25    &    3.5    \\
1563-66L    &    WC8:    &    -284    &    -82    &    -25    &    -25   &    3.5    \\
1583-48L    &    WC8    &    -412    &    -158    &    -30    &    -53   &    2.6    \\
1583-47L    &    WC8    &    -297    &    -107    &    -15    &    -33   &    2.8    \\
1603-75L    &    WC8    &    -370    &    -132    &    ---    &    -37   &    2.8    \\
1675-10L    &    WC8    &    -292    &    -115    &    -19    &    -31   &    2.5    \\
1085-72L    &    WC8-9    &    -56    &    -113    &    -28    &   -25    &    0.5    \\
1054-43L    &    WC9    &    ---    &    -82    &    -70    &    -12   &    ---    \\
1097-71L    &    WC9    &    -48    &    -155    &    -72    &    -34   &    0.3    \\
1106-31L    &    WC9    &    -10    &    -100    &    -61    &    -32   &    0.1    \\
1133-59L    &    WC9    &    -57    &    -155    &    -50    &    -46   &    0.4    \\
1189-110L    &    WC9    &    -61    &    -124    &    -34    &    -31   &    0.5    \\
1245-23L    &    WC9    &    -5    &    -53    &    -55    &    -19   &    0.1    \\
1327-25L    &    WC9    &    -25    &    -56    &    -15    &    -17   &    0.4    \\
1381-20L    &    WC9    &    -11    &    -126    &    -72    &    -37   &    0.1    \\
1477-55L    &    WC9    &    -46    &    -110    &    -37    &    -33   &    0.4    \\
1487-80L    &    WC9    &    -39    &    -88    &    -39    &    -27   &    0.4    \\
1489-36L    &    WC9    &    -5    &    -87    &    -55    &    -34   &    0.1    \\
1522-55L    &    WC9    &    -46    &    -116    &    -35    &    -32   &    0.4    \\
1093-59L    &    WC9+late-type spectrum    &    -81    &    -66    &   -9    &    -18    &    1.2    \\
\enddata
\end{deluxetable}

\begin{deluxetable}{lccccccrrrrrrrr}
\label{tab:tab1}
\tabletypesize{\scriptsize}
\tablecaption{New WC stars organized by subtype\label{tab:WCsubtype}}
\tablewidth{0pt}
\tablehead{
\colhead{Name} &
\colhead{Subtype} &
\colhead{$\Delta$ $m_{He I}$} &
\colhead{$\Delta$ $m_{C IV}$} &
\colhead{$\Delta$ $m_{Br \gamma}$}  &
\colhead{$\Delta$ $m_{He II}$}  \\
\colhead{(1)}  &
\colhead{(2)}  &
\colhead{(3)}  &
\colhead{(4)}  &
\colhead{(5)}  &
\colhead{(6)}  \\
}
\startdata
1077-55L    &    WC6:    &    5.97    &    10.13    &    0.09    &   1.31    \\
1179-129L    &    WC6:    &    11.77    &    15.64    &    0.96    &   1.41    \\
1670-57L    &    WC6:    &    12.54    &    22.18    &    2.37    &   3.13    \\
1669-24L    &    WC6:    &    9.24    &    17.22    &    0.51    &   3.16    \\
1023-63L    &    WC7:    &    ---    &    ---    &    ---    &    ---    \\
1038-22L    &    WC7:    &    4.94    &    10.91    &    2.40    &   3.73    \\
1051-67L    &    WC7:    &    9.54    &    17.62    &    -0.28    &   0.54    \\
1093-140L&        WC7:    &    10.58    &    12.31    &    -0.26    &   0.49    \\
1095-189L&        WC7:    &    6.31    &    15.74    &    0.40    &   0.93    \\
1109-74L    &    WC7:    &    12.38    &    16.60    &    0.82    &   1.29    \\
1168-91L    &    WC7:    &    7.29    &    16.26    &    1.10    &   2.43    \\
1513-111L&        WC7:    &    6.67    &    11.31    &    1.11    &   2.76    \\
1528-15L    &    WC7    &    7.56    &    9.44    &    0.59    &   0.27    \\
1563-89L    &    WC7:    &    6.24    &    12.74    &    -0.29    &   0.58    \\
1567-51L    &    WC7:    &    6.18    &    13.18    &    -0.42    &   1.06    \\
1583-64L    &    WC7:    &    5.59    &    12.10    &    0.40    &   0.93    \\
1657-51L    &    WC7    &    14.70    &    20.14    &    0.26    &   1.83    \\
1652-24L    &    WC7:    &    10.32    &    20.71    &    2.94    &   6.89    \\
1675-17L    &    WC7:    &    14.47    &    21.24    &    2.21    &   5.06    \\
1085-69L    &    WC8    &    4.24    &    10.78    &    -0.07    &   0.31    \\
1085-83L    &    WC8:    &    4.74    &    14.63    &    0.69    &   1.93    \\
1093-138L&        WC8:    &    9.50    &    18.00    &    0.47    &   0.87    \\
1091-46L    &    WC8    &    3.15    &    5.54    &    0.56    &   0.88    \\
1097-34L    &    WC8    &    9.14    &    9.87    &    1.29    &   2.01    \\
1105-76L    &    WC8    &    8.44    &    17.78    &    2.27    &   3.90    \\
1181-82L    &    WC8    &    11.68    &    17.32    &    1.40    &   1.61    \\
1181-81L    &    WC8    &    10.64    &    15.46    &    2.85    &   3.96    \\
1269-166L&        WC8    &    7.95    &    9.70    &    3.79    &   3.58    \\
1395-86L    &    WC8    &    4.98    &    7.32    &    0.71    &   0.96    \\
1434-43L    &    WC8    &    6.60    &    17.56    &    0.58    &   1.75    \\
1463-7L    &    WC8    &    4.49    &    7.76    &    2.44    &   2.52    \\
1493-9L    &    WC8    &    3.80    &    9.56    &    2.53    &   1.50    \\
1495-32L    &    WC8    &    4.80    &    14.81    &    3.06    &   3.04    \\
1527-13L    &    WC8    &    4.45    &    6.86    &    1.07    &   3.27    \\
1551-19L    &    WC8:    &    4.45    &    12.06    &    0.28    &   1.29    \\
1563-66L    &    WC8:    &    6.73    &    10.46    &    0.84    &   1.11    \\
1583-48L    &    WC8    &    5.96    &    14.28    &    1.39    &   2.69    \\
1583-47L    &    WC8    &    5.10    &    11.92    &    0.38    &   1.18    \\
1603-75L    &    WC8    &    5.16    &    16.28    &    1.08    &   1.47    \\
1675-10L    &    WC8    &    5.15    &    12.04    &    -0.79    &   0.27    \\
1085-72L    &    WC8-9    &    4.01    &    7.13    &    1.67    &   0.96    \\
1054-43L    &    WC9    &    8.43    &    7.42    &    3.38    &   1.08    \\
1097-71L    &    WC9    &    11.44    &    6.79    &    1.99    &   1.27    \\
1106-31L    &    WC9    &    18.03    &    5.72    &    8.08    &   4.25    \\
1133-59L    &    WC9    &    4.95    &    6.50    &    2.22    &   1.51    \\
1189-110L&        WC9    &    5.55    &    5.48    &    1.83    &   1.60    \\
1245-23L    &    WC9    &    13.75    &    1.65    &    3.41    &   0.69    \\
1327-25L    &    WC9    &    ---    &    ---    &    ---    &    ---    \\
1381-20L    &    WC9        &---    &    ---    &    ---    &    ---    \\
1477-55L    &    WC9    &    6.63    &    5.40    &    1.14    &   0.87    \\
1487-80L    &    WC9    &    8.99    &    5.61    &    0.99    &   0.69    \\
1489-36L    &    WC9    &    7.17    &    1.37    &    3.64    &   0.42    \\
1522-55L    &    WC9    &    7.07    &    4.64    &    0.74    &   0.83    \\
1093-59L    &    WC9    &    4.83    &    5.64    &    -0.51    &   -0.04    \\
\hline
\enddata
\tablecomments{$\Delta$ Magnitudes calculated from the narrow band images collected from our Galactic Plane Survey}

\end{deluxetable}

\begin{deluxetable}{lccccccrrrrrrrr}
\label{tab:tab1}
\tabletypesize{\scriptsize}
\tablecaption{New WN stars organized by subtype\label{tab:WNsubtype}}
\tablewidth{0pt}
\tablehead{
\colhead{Name} &
\colhead{Subtype} &
\colhead{$\Delta$$m_{He I}$} &
\colhead{$\Delta$$m_{C IV}$} &
\colhead{$\Delta$$m_{Br \gamma}$}  &
\colhead{$\Delta$$m_{He II}$}  \\
\colhead{(1)}  &
\colhead{(2)}  &
\colhead{(3)}  &
\colhead{(4)}  &
\colhead{(5)}  &
\colhead{(6)}  \\
}
\startdata
1322-220L    &    WN5    &    -0.79    &    -0.86    &    3.32    &   6.17    \\
1536-180L    &    WN5    &    -1.40    &    -1.19    &    3.44    &   9.03    \\
1097-156L    &    WN6:    &    1.34    &    0.01    &    3.10    &   7.17    \\
1115-197L    &    WN6    &    -0.07    &    -0.91    &    3.34    &   4.99    \\
1138-133L    &    WN6    &    -0.57    &    -1.30    &    3.45    &   11.53    \\
1342-208L    &    WN6    &    -1.16    &    -0.84    &    5.60    &   11.58    \\
1650-96L    &    WN6    &    2.87    &    0.35    &    4.42    &   3.26    \\
1698-70L    &    WN6    &    -2.34    &    -1.89    &    6.82    &   14.20    \\
1181-211L    &    WN7:    &    0.70    &    -0.17    &    3.06    &   4.12    \\
1483-212L    &    WN7:    &    0.85    &    -0.74    &    4.25    &   4.43    \\
1487-212L    &    WN7    &    0.60    &    0.28    &    3.45    &   4.28    \\
1503-160L    &    WN7    &    2.11    &    0.04    &    5.70    &   6.26    \\
1517-138L    &    WN7    &    1.55    &    -0.74    &    3.89    &   5.30    \\
1042-25L    &    WN8    &    4.22    &    0.11    &    5.04    &   2.38    \\
1275-184L    &    WN8    &    5.47    &    -0.50    &    4.30    &   2.27    \\
1431-34L    &    WN8    &    3.06    &    -0.11    &    5.01    &   3.36    \\
1093-140LB    &    WN9    &    ---    &    ---    &    ---    &    ---    \\
\hline
\enddata
\tablecomments{$\Delta$ Magnitudes calculated from the narrow band images collected from our Galactic Plane Survey}

\end{deluxetable}
 
\clearpage

\begin{figure*}[!ht]
\begin{center}
\epsscale{1.0}
\plotone{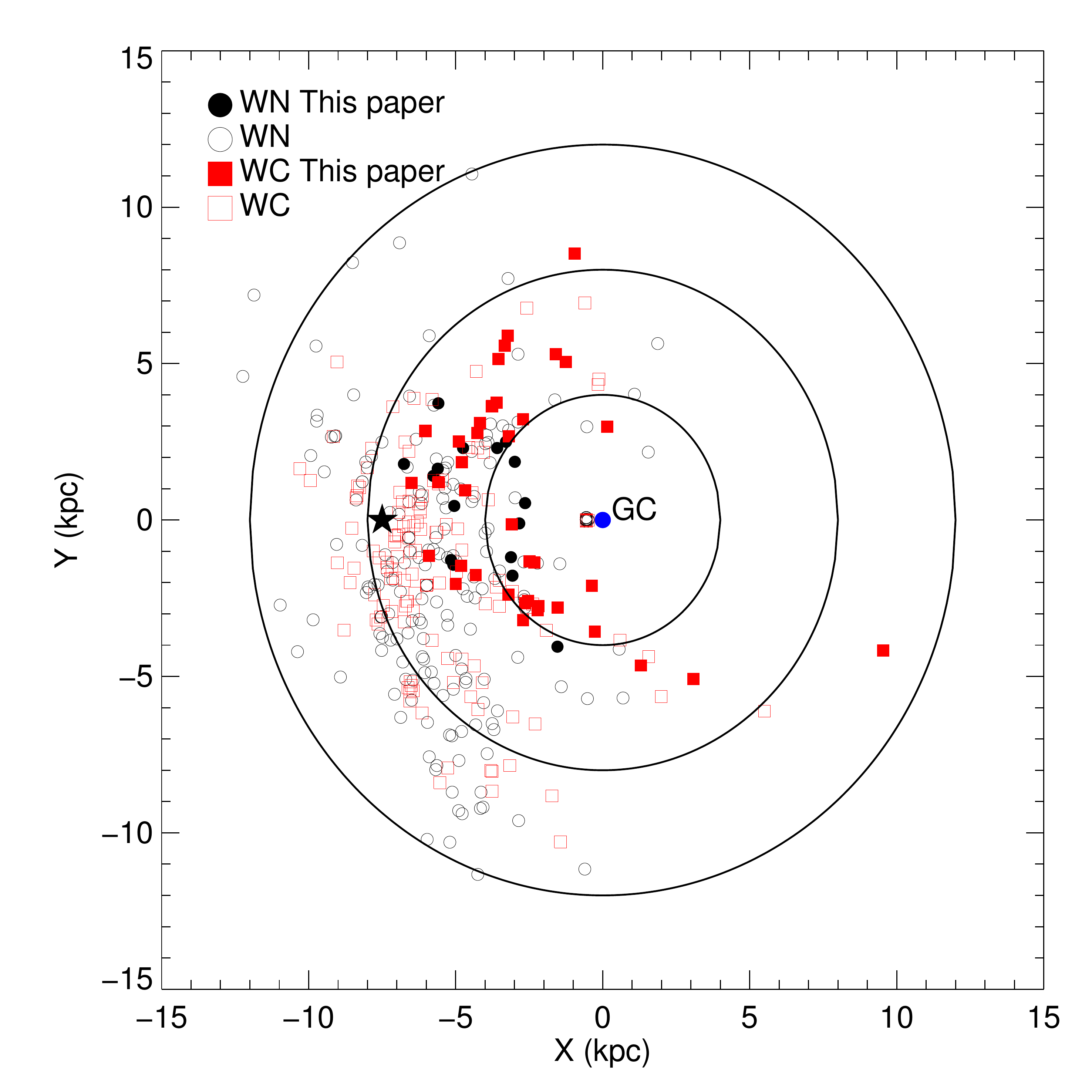}
\end{center}
\caption{Galactic distribution of known and new WR stars with estimated distances projected on the plane.  New WC and WN stars are filled boxes and circles respectively while known WC and WN stars are unfilled.  The Galactic Center (GC) is labeled as is the position of the Sun (black five-pointed star).  Circles of radius 4, 8, and 12 kpc are overplotted.}
\label{Galaxyplot}
\end{figure*}

In Figure \ref{Galaxyplot} our 71 new WR stars (in bold) have been plotted together with 321 previously-known WRs onto the plane of the Galaxy. The distances to the other stars with established distances were taken from discovery papers and the seventh WR catalog (\citealt{had07}, \citealt{mau09,mau10,mau10b,mau10c,mau11}, \citealt{sha99}, \citealt{sha09}, \citealt{vdh01}). The Galactic center is labeled, and circles of radius 4, 8, and 12 kpc are plotted. The Sun is indicated with a five-pointed star. The 71 new WR stars are located at significantly larger heliocentric distances than most other known stars. We also find a few new stars without optical counterparts within just a few kpc of our Sun, reinforcing the necessity of WR surveys in the near infrared. \citet{con90}, along with the more recent reanalysis in \citet{had07}, maintain that WR stars trace the spiral structure of the Galaxy. One arm may be seen along roughly the 8 kpc radius, and an inner arm can perhaps begin to be traced along the inner 4kpc radius. However, the distance error bars are not trivial, so that firm conclusions about the utility of WR stars as spiral tracers cannot yet be drawn.

\pagebreak

\pagebreak

The spectra of our new WR stars are shown in Figures 2-10 . Overall, few early subtypes of either sequence (WN or WC) were seen, again as expected in the inner Galaxy for higher-than-solar Z. Selection bias in favor of late subtypes is unlikely to be operating, since the early subtypes are the easiest WR stars to find. We are confident that they would have been found, given their strong HeII lines (WNE stars) and strong HeII and CIV lines (WCE stars). Note the contrast with the outer Galaxy, where earlier types abound, as in M33 \citep{neu11}, the LMC and especially the SMC, where Z is progressively smaller.  The physical reason for this is now recognized to be due to Z-related opacity effects. For lower Z, mass loss rates are lower AND one can see deeper into hotter layers of the wind.  Thus, what might be a WCL star with Z at twice the solar value would instead be seen as a WCE star in the LMC (where Z can be half solar).  Among the 24 WC/WO stars in the LMC, 23 are WC4 and one is WO4; nothing cooler is seen.  Only two WC7 stars are known in M33 \citep{neu11}; all others are of earlier types. In the Local Group, WC9 stars are only found in the inner Galaxy and possibly in the metal-richest parts of M31.

\clearpage

\begin{figure*}[!ht]
\begin{center}
\epsscale{1.0}
\plotone{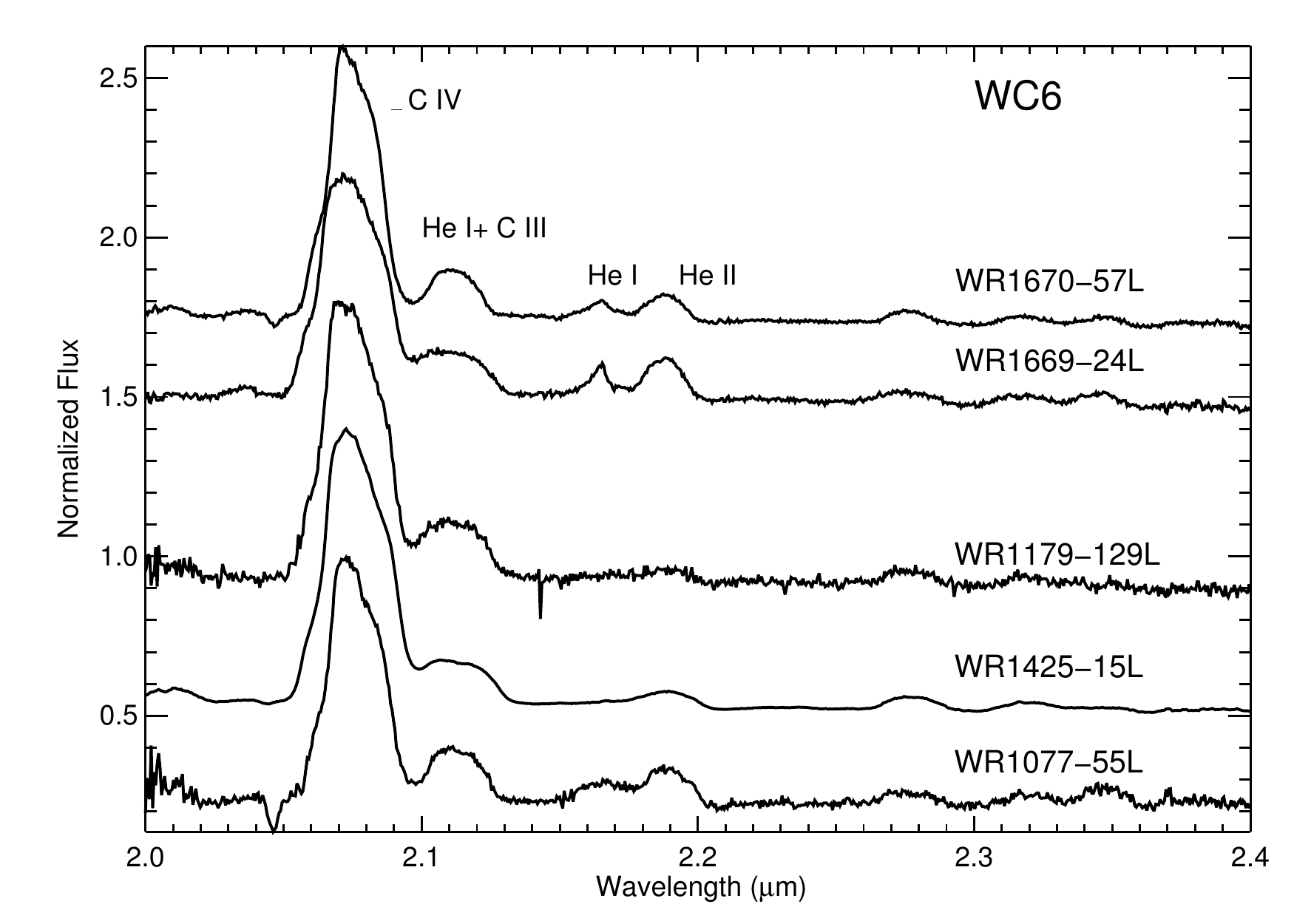}
\end{center}
\caption{All new WC6 objects classified in this work as well as one previously identified object.}
\label{fig:WC6}
\end{figure*}

\begin{figure*}[!ht]
\begin{center}
\epsscale{1.0}
\plotone{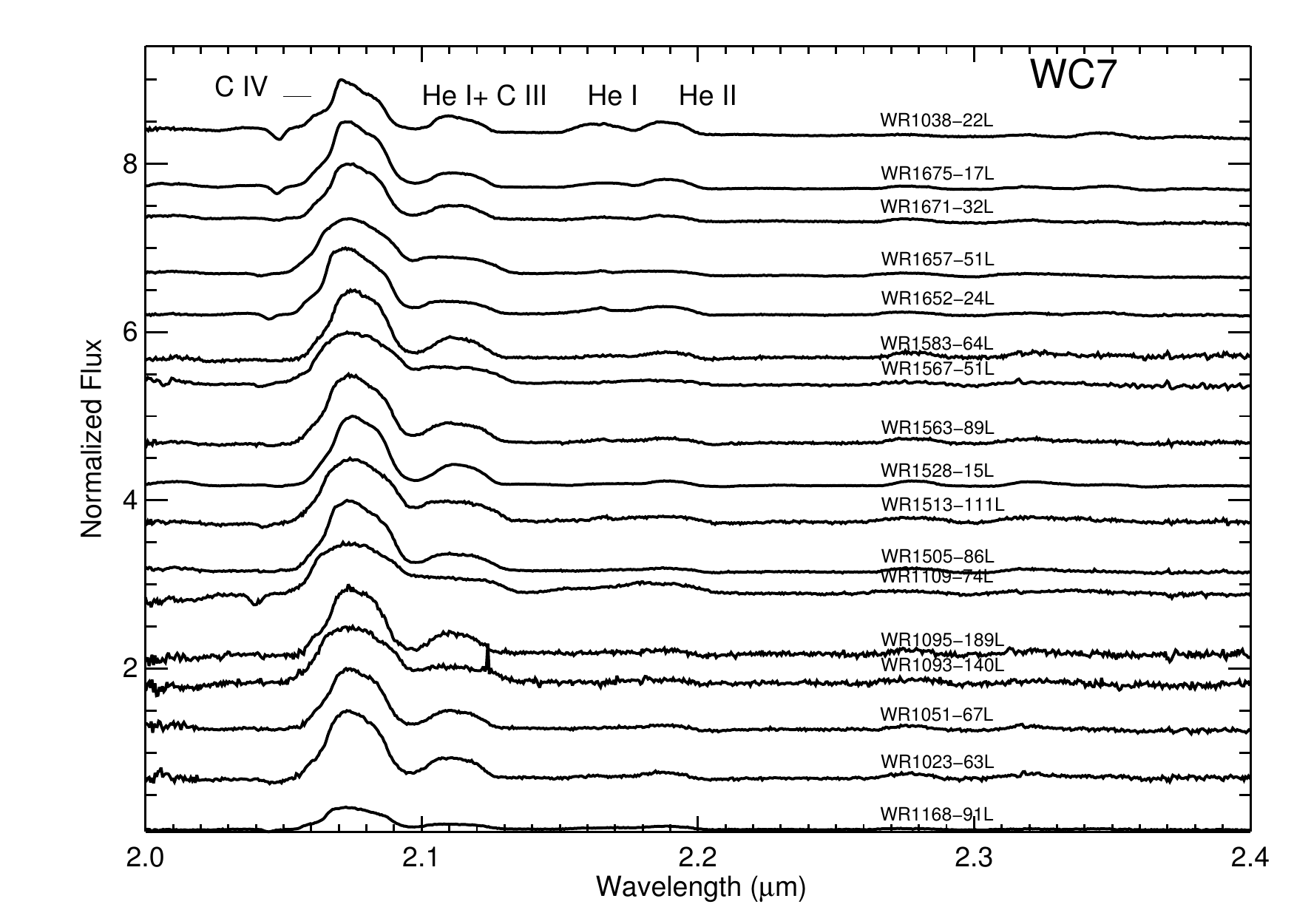}
\end{center}
\caption{All new WC7 objects classified in this work as well as two previously identified objects.}
\label{fig:WC7}
\end{figure*}

\begin{figure*}[!ht]
\begin{center}
\epsscale{1.0}
\plotone{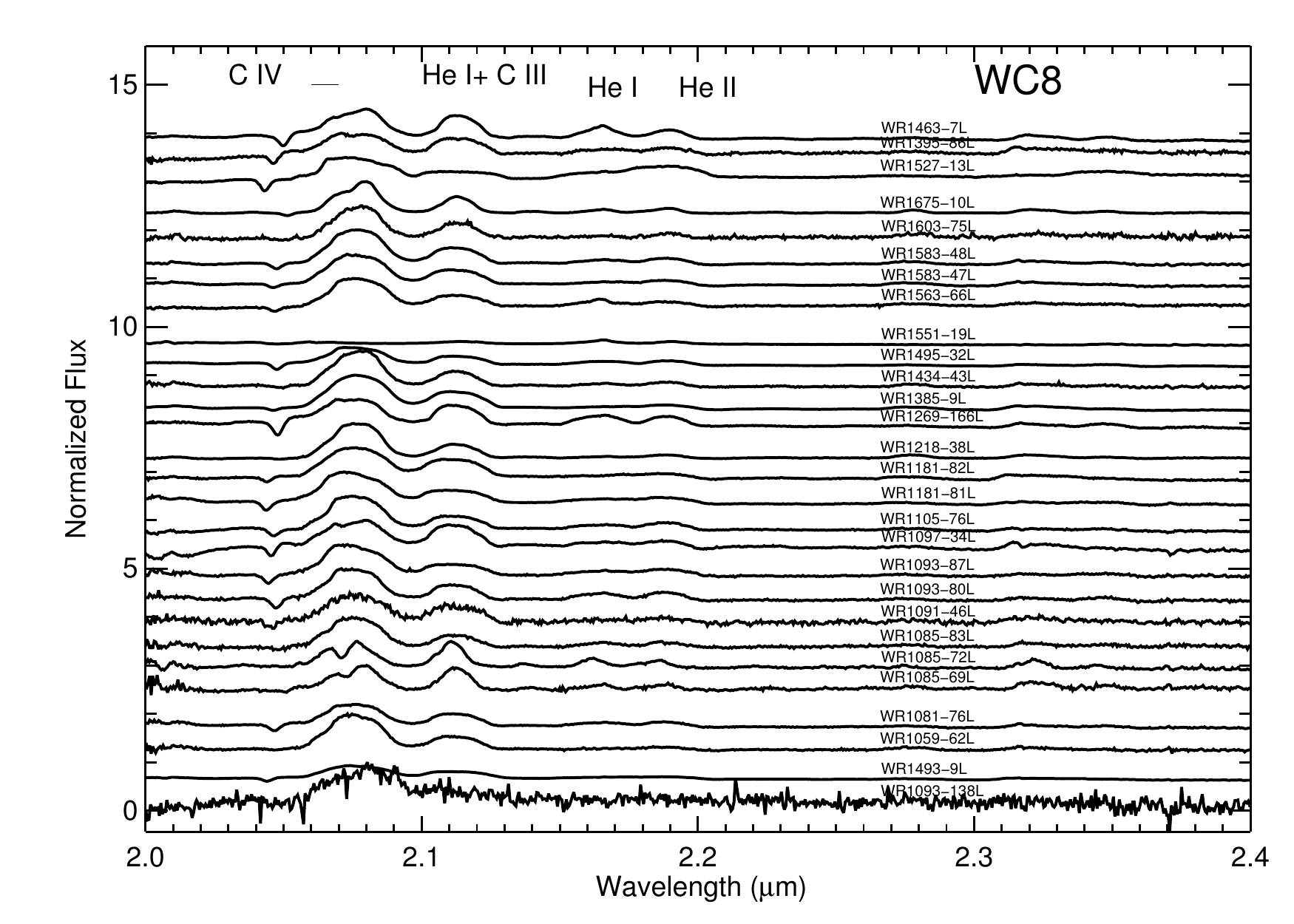}
\end{center}
\caption{All new WC8 objects classified in this work as well as six previously identified objects.}
\label{fig:WC8}
\end{figure*}

\begin{figure*}[!ht]
\begin{center}
\epsscale{1.0}
\plotone{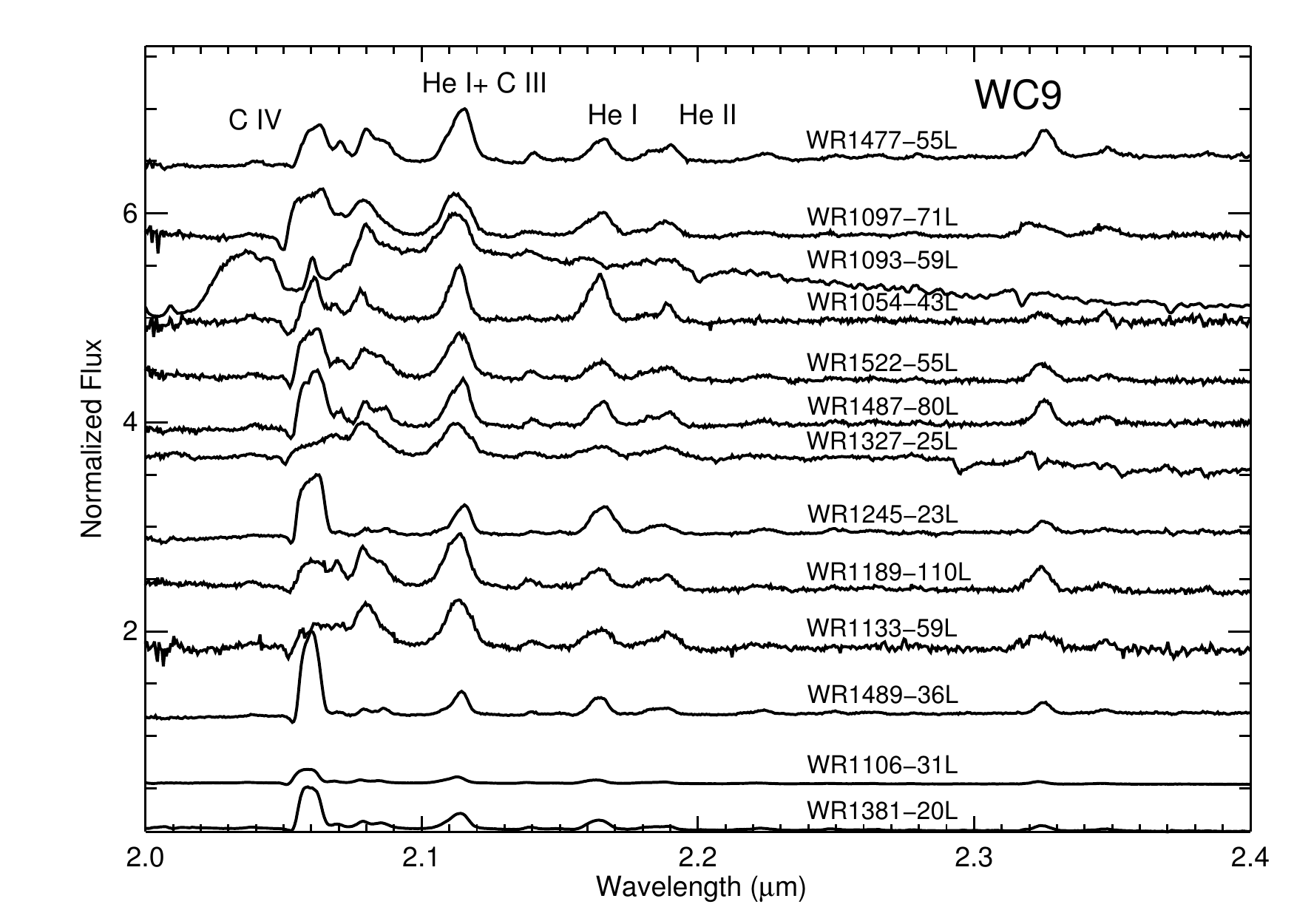}
\end{center}
\caption{All new WC9 objects classified in this work.}
\label{fig:WC9}
\end{figure*}

\begin{figure*}[!ht]
\begin{center}
\epsscale{1.0}
\plotone{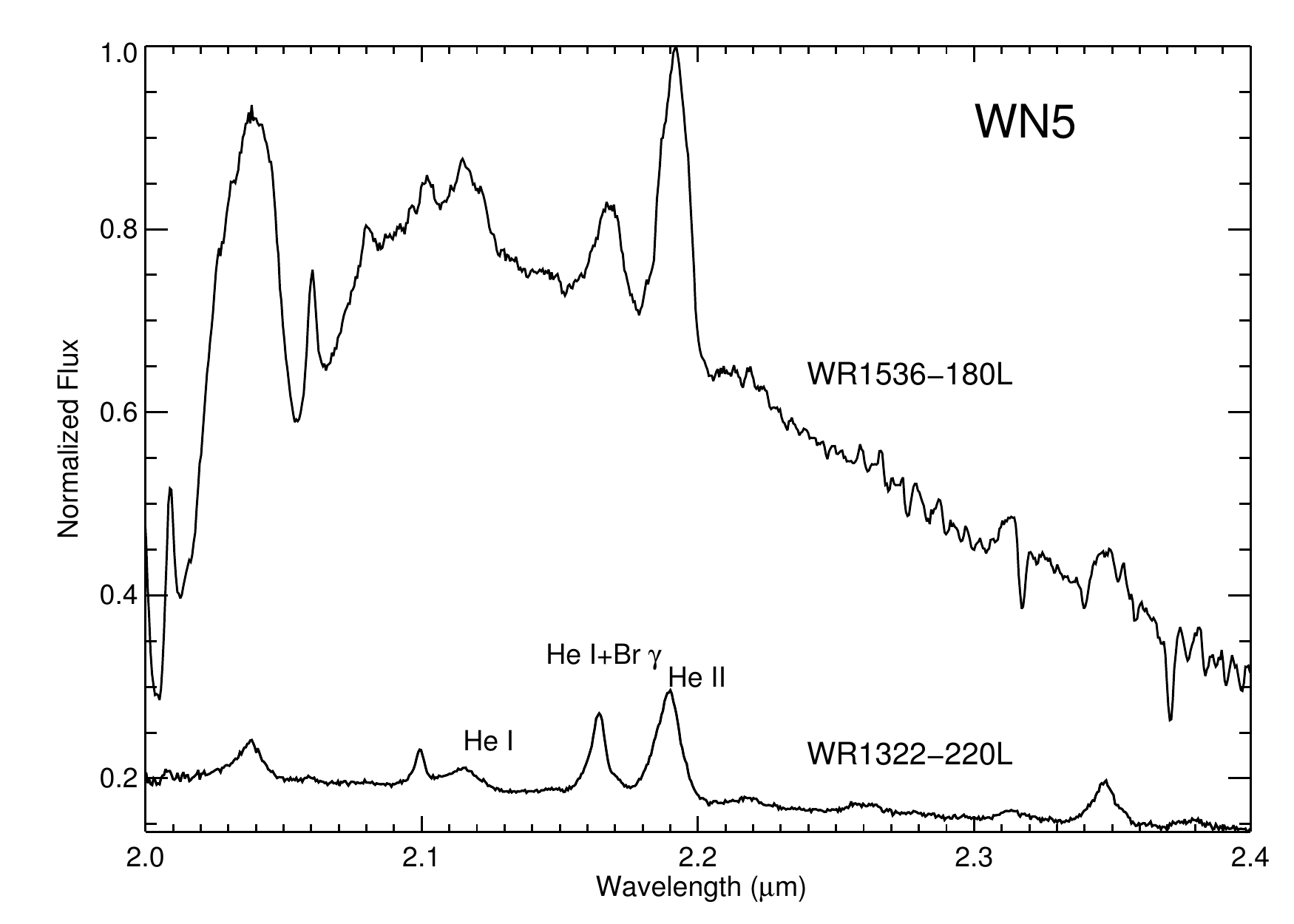}
\end{center}
\caption{All new WN5 objects classified in this work.}
\label{fig:WN5}
\end{figure*}

\begin{figure*}[!ht]
\begin{center}
\epsscale{1.0}
\plotone{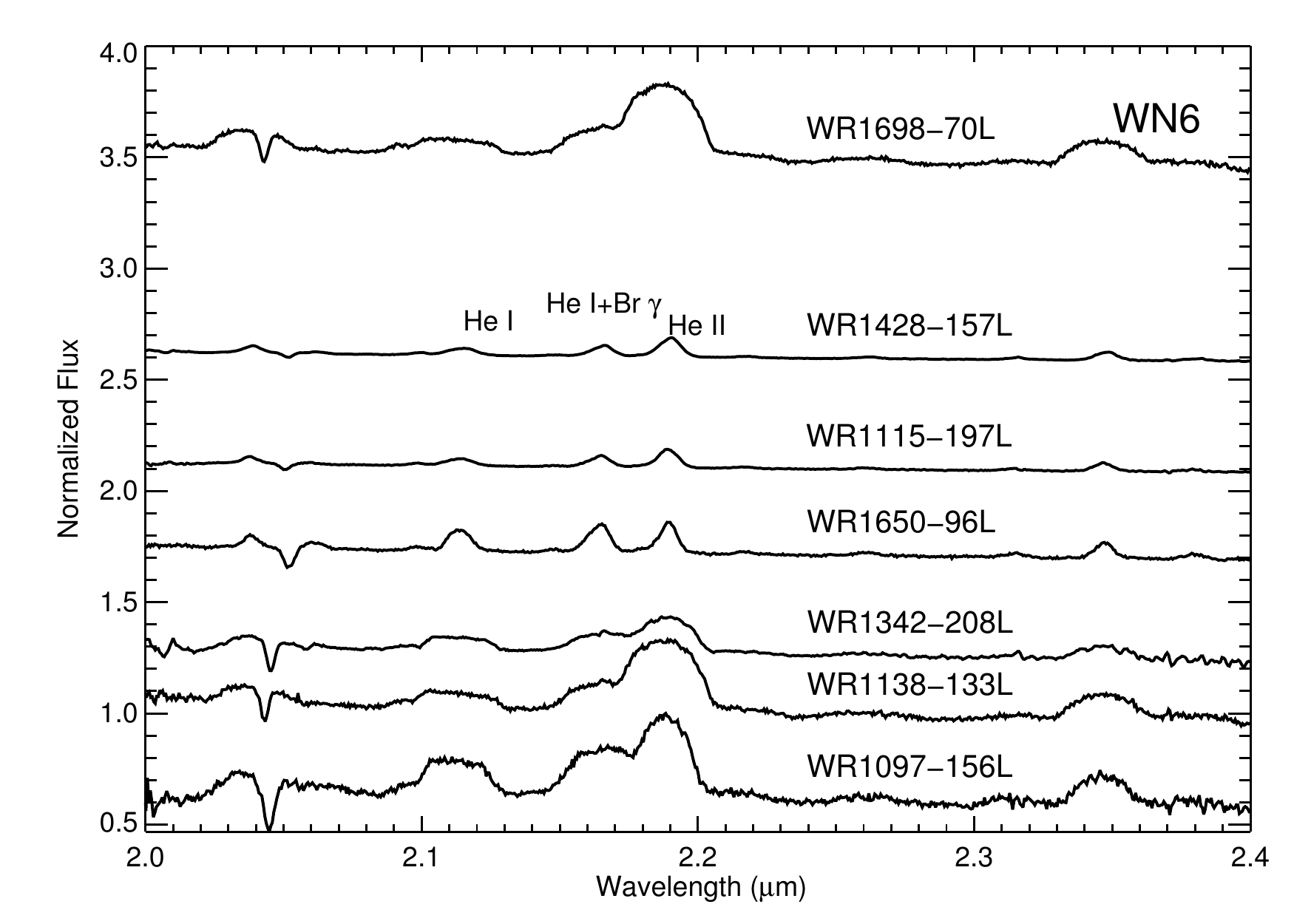}
\end{center}
\caption{All new WN6 objects classified in this work as well as one previously identified object.}
\label{fig:WN6}
\end{figure*}

\begin{figure*}[!ht]
\begin{center}
\epsscale{1.0}
\plotone{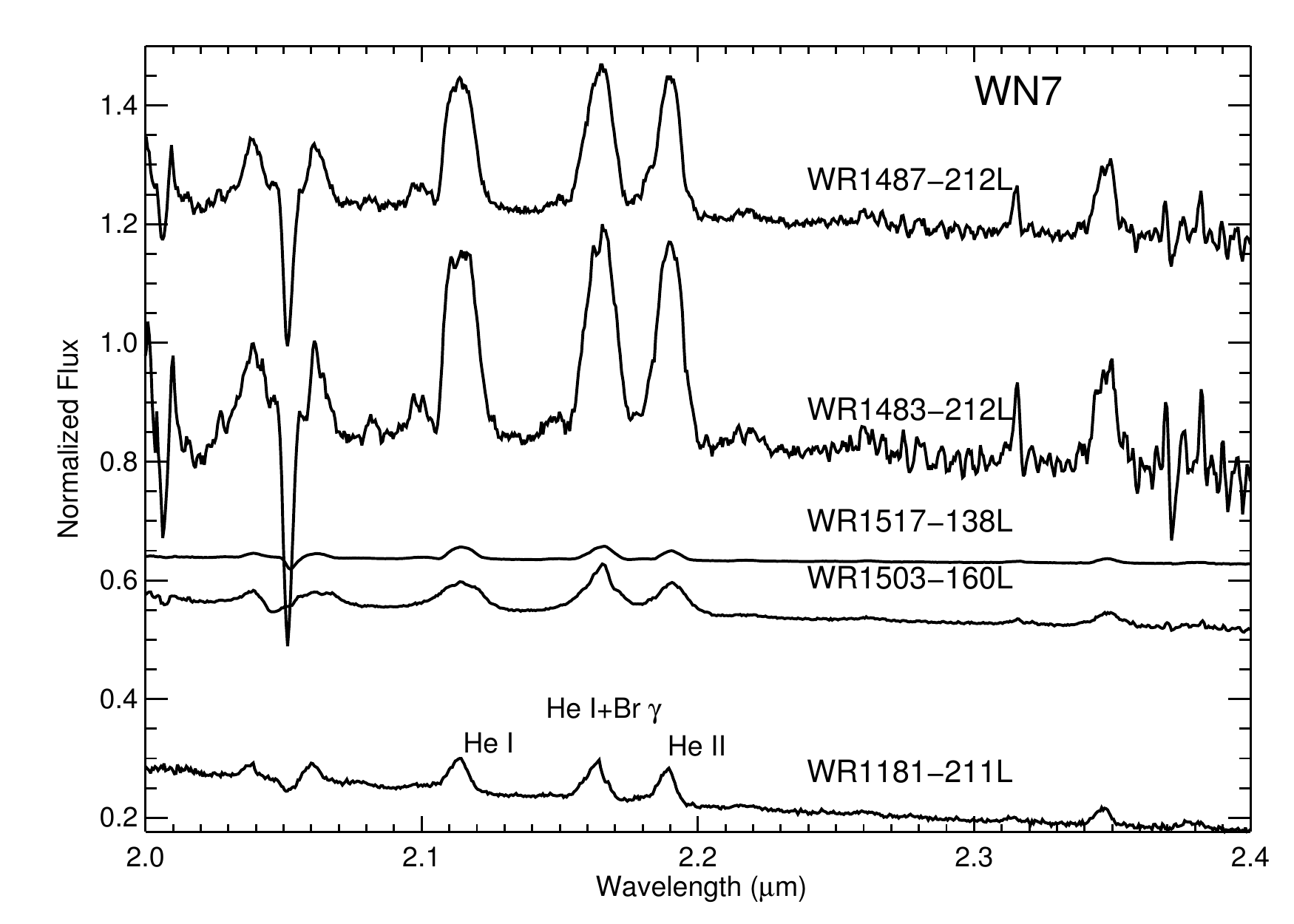}
\end{center}
\caption{All new WN7 objects classified in this work.}
\label{fig:WN7}
\end{figure*}

\begin{figure*}[!ht]
\begin{center}
\epsscale{1.0}
\plotone{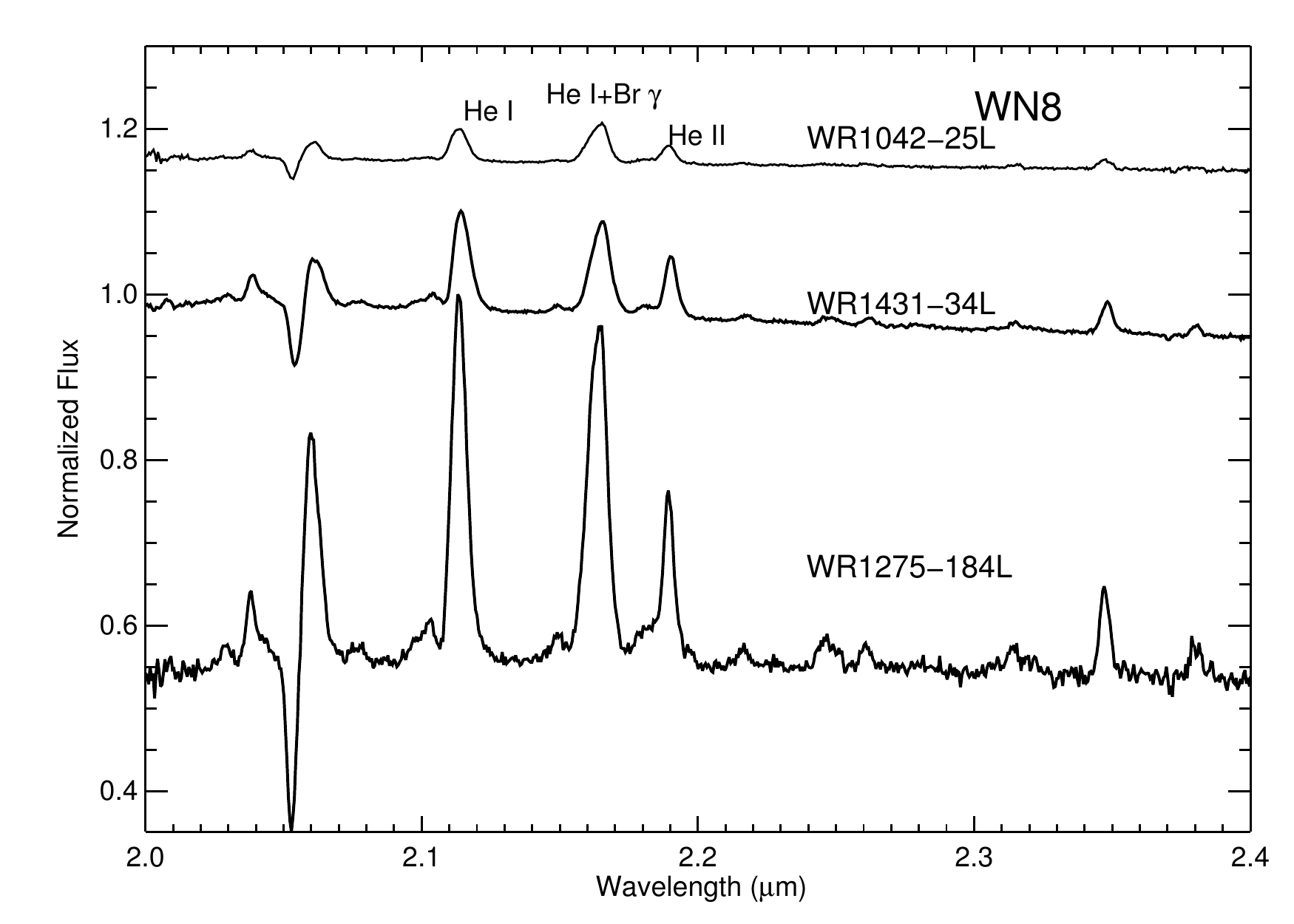}
\end{center}
\caption{All new WN8 objects classified in this work.}
\label{fig:WN8}
\end{figure*}

\begin{figure*}[!ht]
\begin{center}
\epsscale{1.0}
\plotone{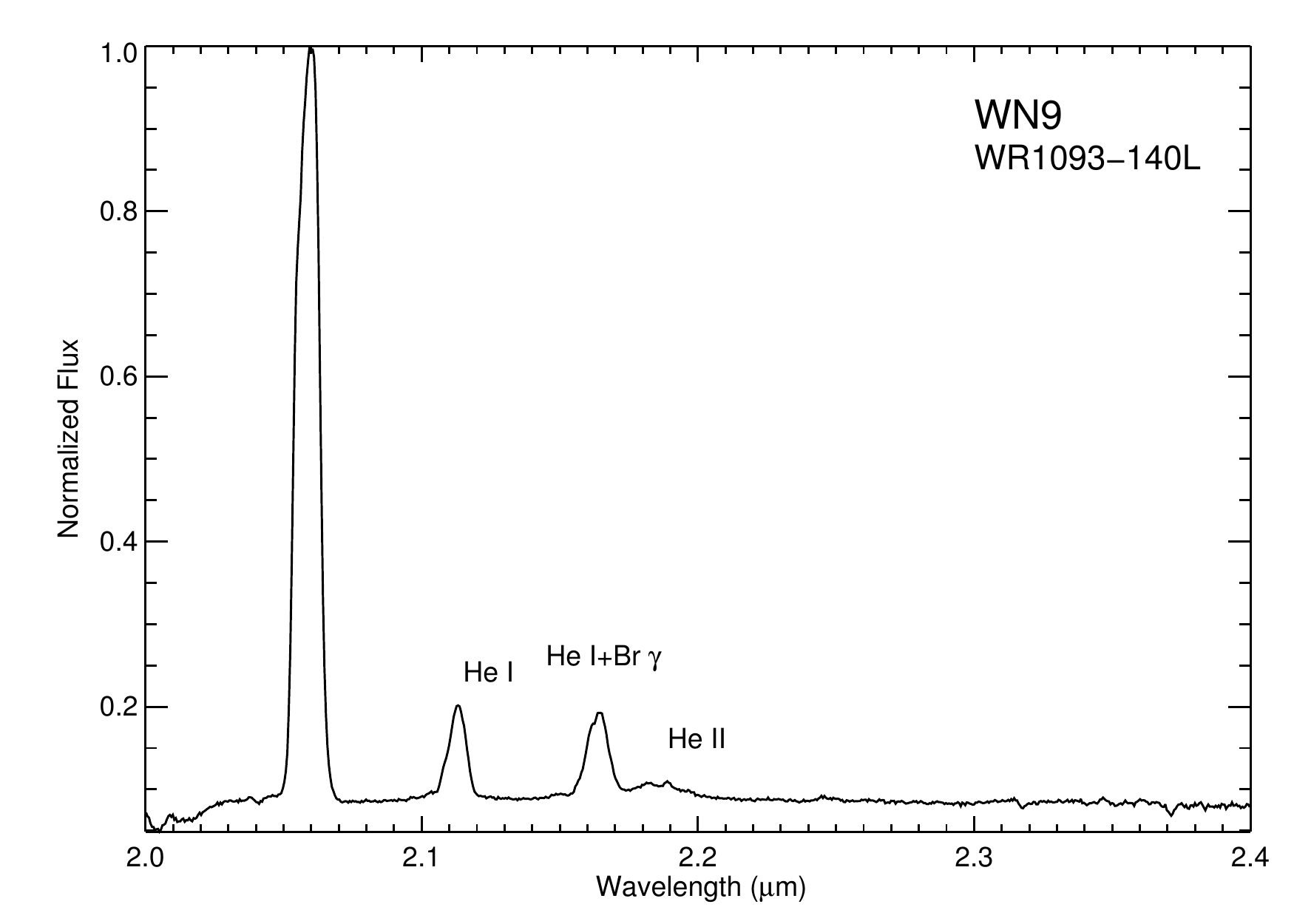}
\end{center}
\caption{The new WN9 object classified in this work.}
\label{fig:WN9}
\end{figure*}

\clearpage

Some types may not have been found: WO (since our filters did not select for O lines) and extreme WC9d stars. Lines are severely veiled by continuum dust emission in these cooler WC types, which often show IR excesses from heated dust being formed in wind-collisions with an orbiting companion.  The latter might best be discovered in broadband near IR + Mid-IR surveys  of the type described by \citet{mau09} and \citet{mau11}.

\section{Duds}

About 43\% our WR candidates turned out not to be WR stars. All of the "duds" resembled one of the four examples we show in Figure 11. In each case the flux of these stars is greater in one or more of our narrowband filters than in
one or both of the continuum filters described in Paper I. The cause is not an emission line, but usually absorption in a continuum filter, or a steeply varying spectrum which also mimicked emission. Using broadband J,H,K plus mid-IR photometry to further filter our narrowband candidates may help us avoid almost all of these duds in the future. This is because the broad color space of \citet{mau11} returns about 95\% early-type emission-line stars. Thus the combination of broadband-IR PLUS narrowband 

\clearpage

\begin{figure}
\centering
\begin{tabular}{cc}
\epsfig{file=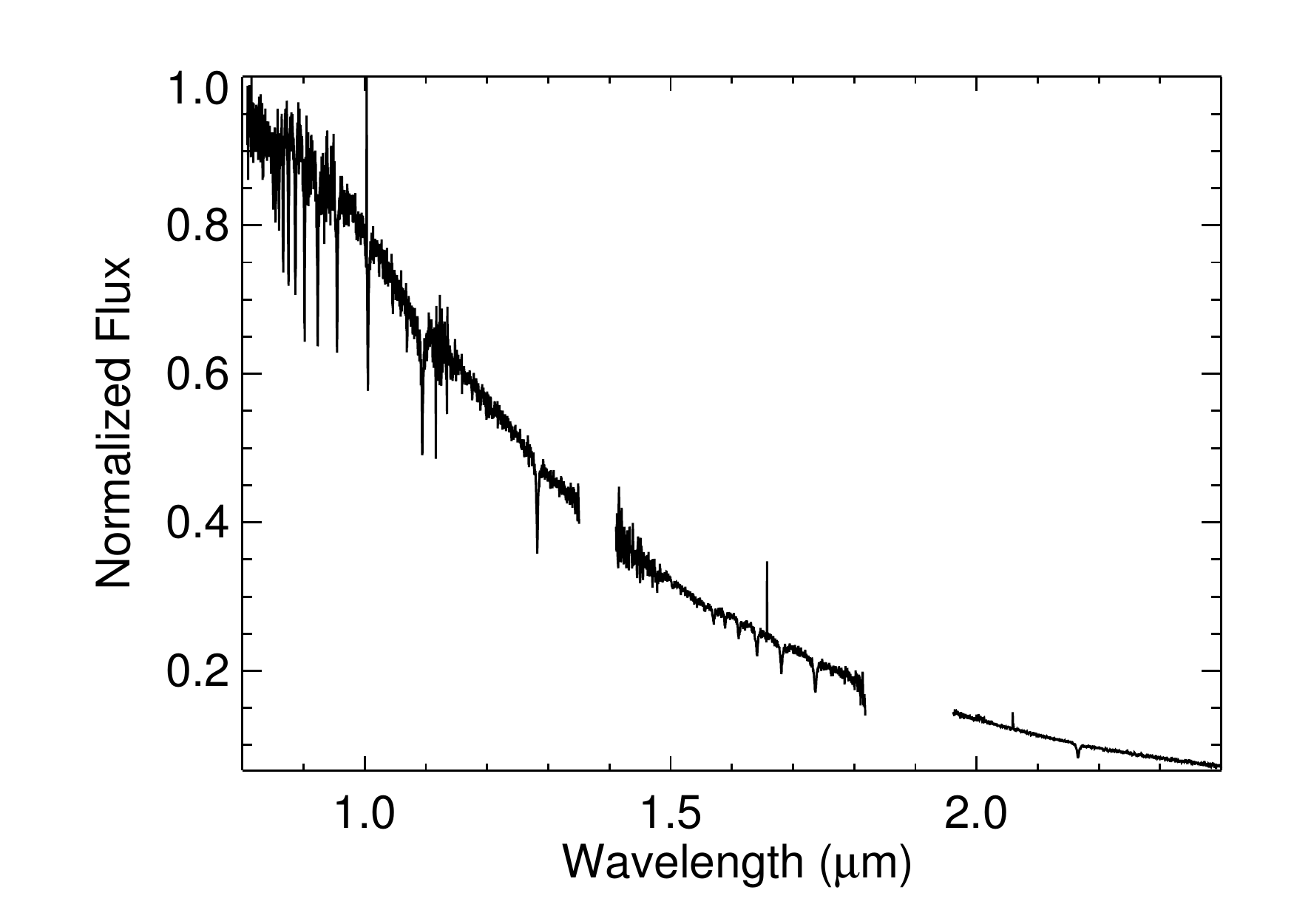,width=0.5\linewidth,clip=} &
\epsfig{file=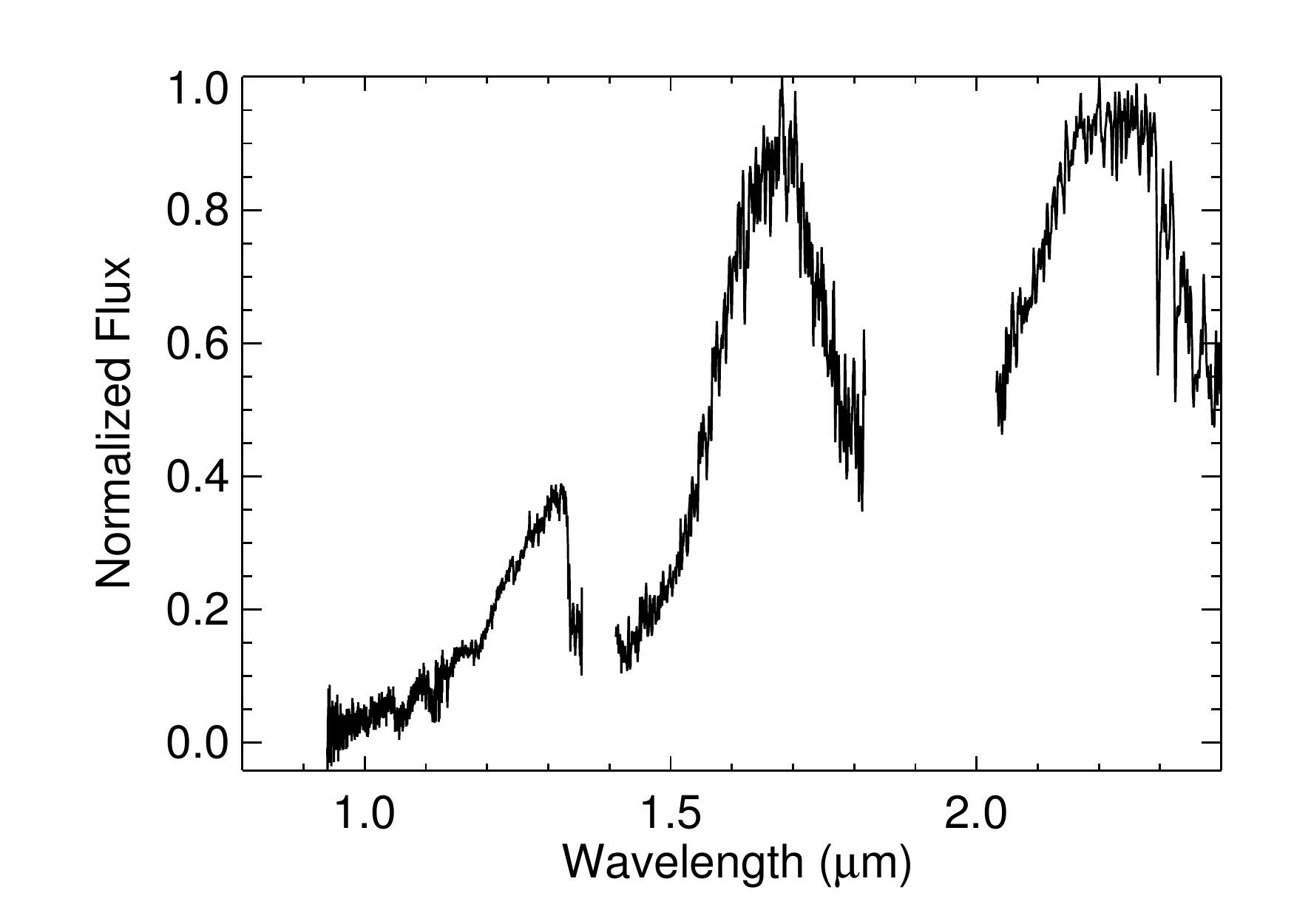,width=0.5\linewidth,clip=} \\
\epsfig{file=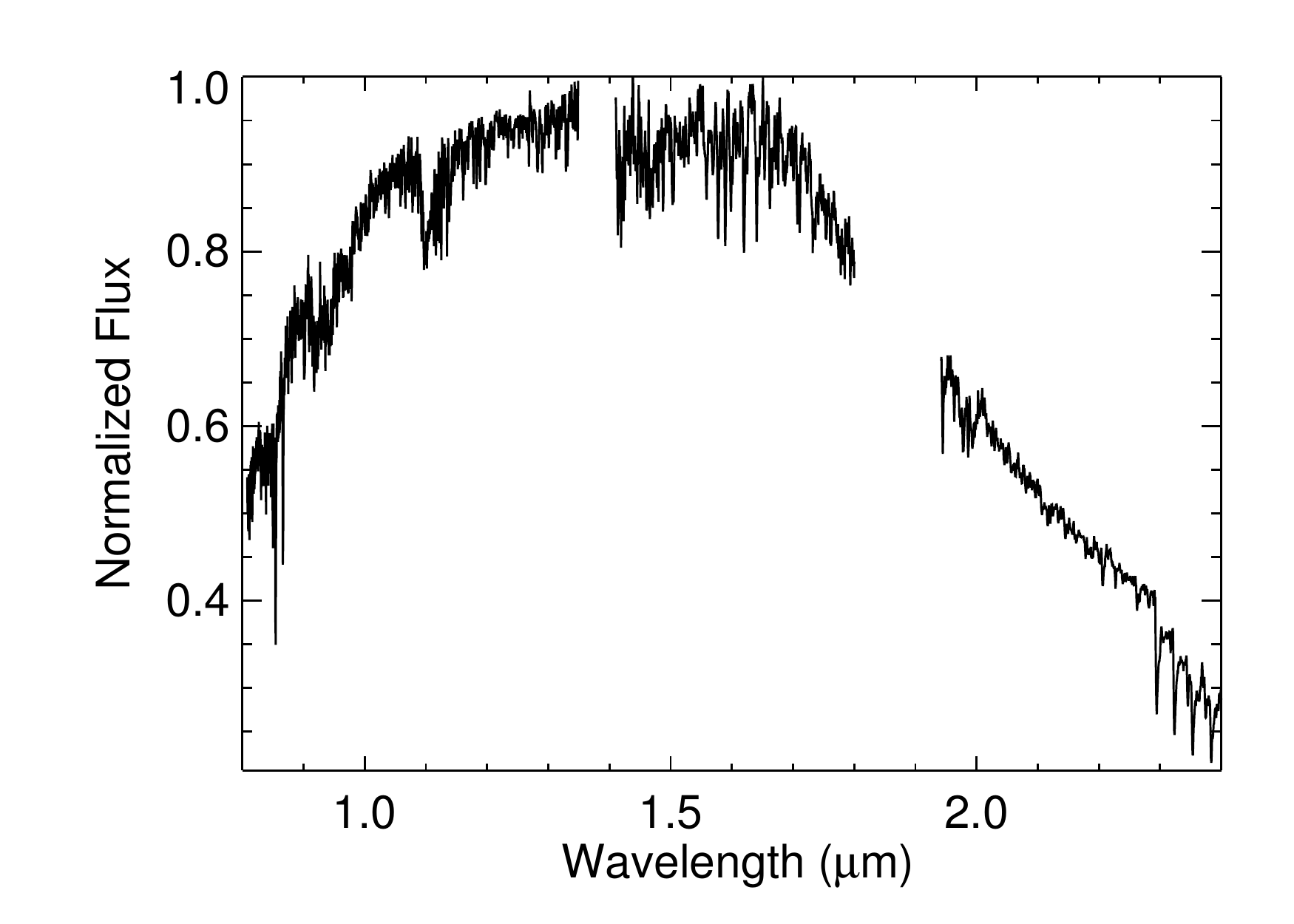,width=0.5\linewidth,clip=} &
\epsfig{file=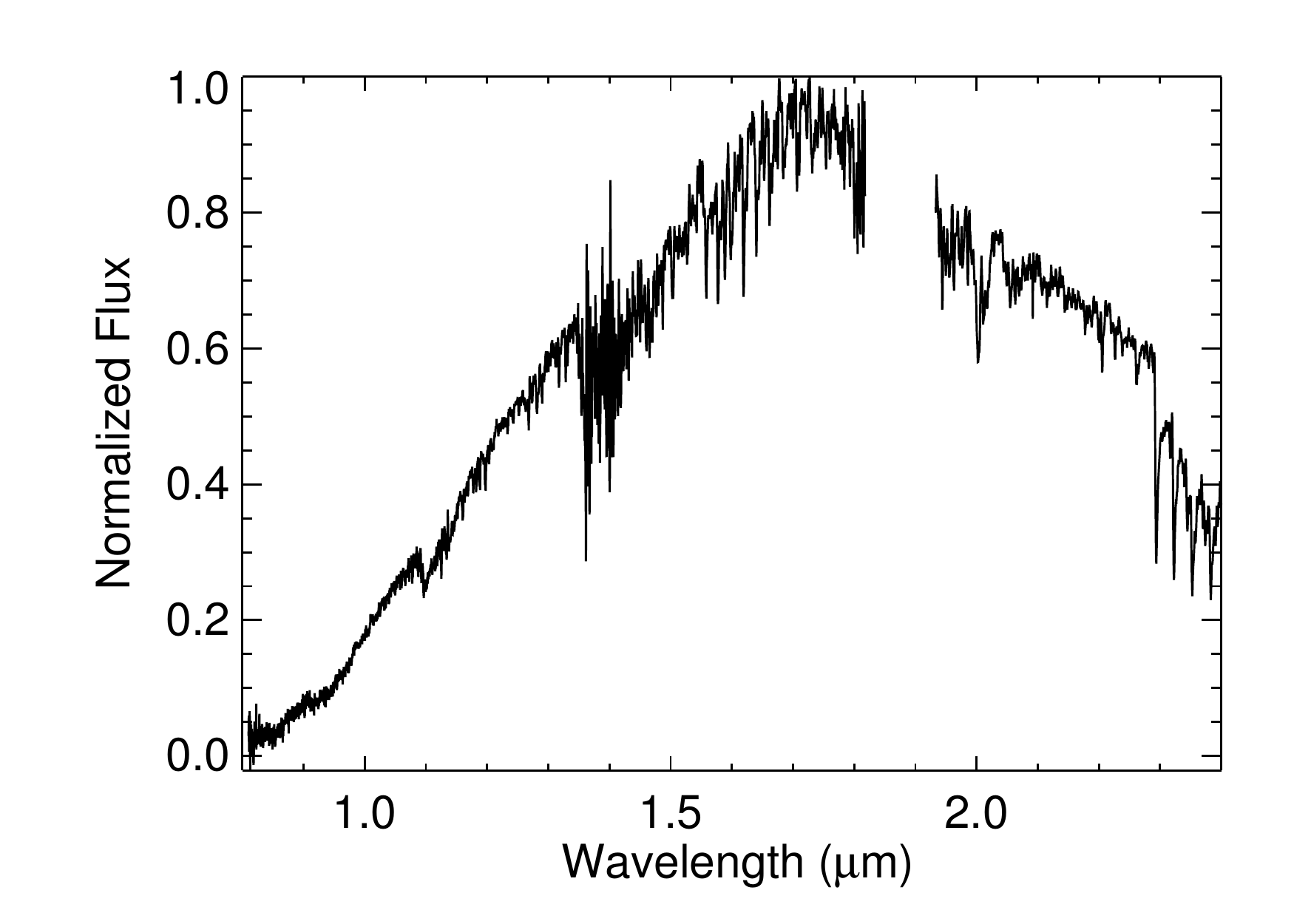,width=0.5\linewidth,clip=}
\end{tabular}
\caption{Four typical examples of objects examined in this work which did not turn out to be Wolf-Rayet stars.  The upper left is a hot F or G type-star, while the three subsequent "duds" are most likely reddened early to late-type M giant stars. }
\label{fig:Duds}
\end{figure}

\clearpage

\section{Three Noteworthy stars}

Two of our new WR stars, 1583-48L and 1583-47L are separated by only 8 arcsec on the sky; both are of subtype WC8, and it is apparent from their finder charts (Figure 12) that they belong to a small, compact cluster.

We also note our new WC7: star 1675-17L, which is seen to have extremely bright arcs of gas emitting predominantly in the lines of HeI and Br-gamma.  

\section{Completeness, Success Rate and Complementarity with IR-color Surveys}

Neither Paper I (41 new WR stars) nor the present paper are complete samples of Wolf-Rayet stars. We reported these 112 new WR stars because they are exceedingly rare, and interesting as potential type Ib and Ic supernovae. They represent our increasingly successful tests of successive generations of image processing pipelines. As described in Paper I, our database of over 77,000 narrowband infrared images is far too vast to analyze in any other than fully automated fashion. The 83 WR stars successfully picked out by our present methodology (including 71 new stars from 146 candidates) demonstrates that 57\% our candidates are bona fide WR stars. This is very encouraging, as infrared spectrographs are much less common than visible-light spectrographs (and of course all telescope time must be used with maximum efficiency). It is important to emphasize that we are reporting mostly WC stars because they are by far the strongest emission-line candidates, and we did not have enough telescope time to do a complete survey.

After this paper was completed we became aware of an astroph paper (now published as \citet{mau11}), which reported 60 new WR star discoveries via infrared color selection. 17 of those new WR stars were also found in the present work, and are amongst the 71 new WR stars reported in this paper. We regard the surveys as complementary. It is certainly correct that the number ratio of WC/WN in our study (54/17) is very different from that found by Mauerhan et al (22/38). Our search area includes a part of the galaxy closer to the Galactic center (where more WC are expected) than Mauerhan et al appear to have searched; and we have not yet spectrographically checked the area l = 284 to 313 degrees which Mauerhan et al did search, and where more WN are expected. This decreases the difference between our results, but only slightly. More important is the fact that WC stars are such powerful emission-line sources that they are the first candidates we have checked. This explains why we find so many more WC than WN stars. We have not yet had enough telescope time to do an area-limited, magnitude limited, equivalent-width limited survey in all our emission-line filters. Thus comparisons between the color-selected and narrowband-selected methods are still premature.
 
\section{Finder Charts} \label{finders}
\begin{figure*}[htbp]
\begin{centering}
\epsscale{.8}
\includegraphics[width=1.1\hsize,angle=90]{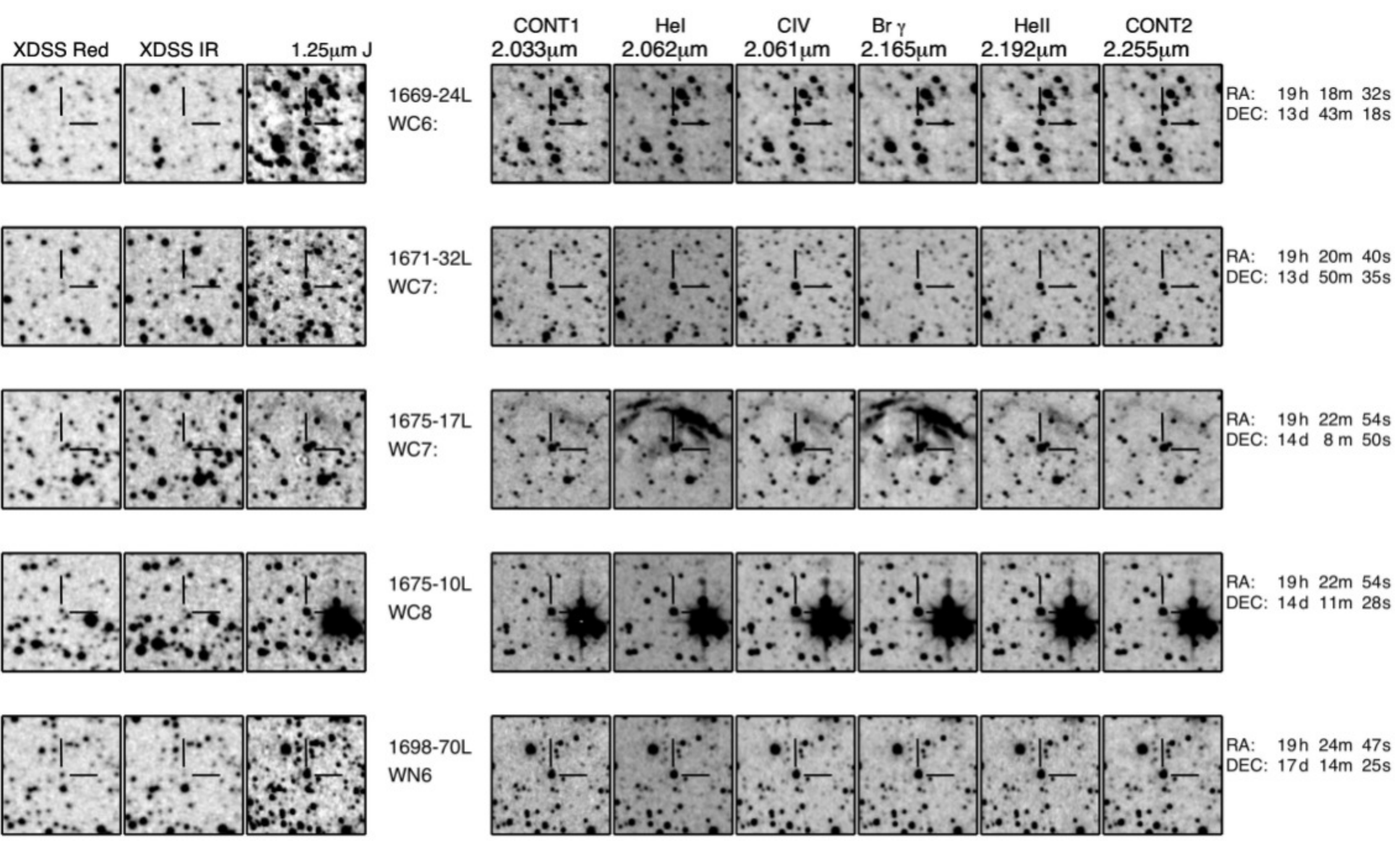}
\end{centering}
\caption{Finder charts for WC and WN stars observed with SpeX.} 
\label{fig:finder1}
\end{figure*}

\begin{figure*}[htbp]
\begin{centering}
\epsscale{.8}
\includegraphics[width=1.1\hsize,angle=90]{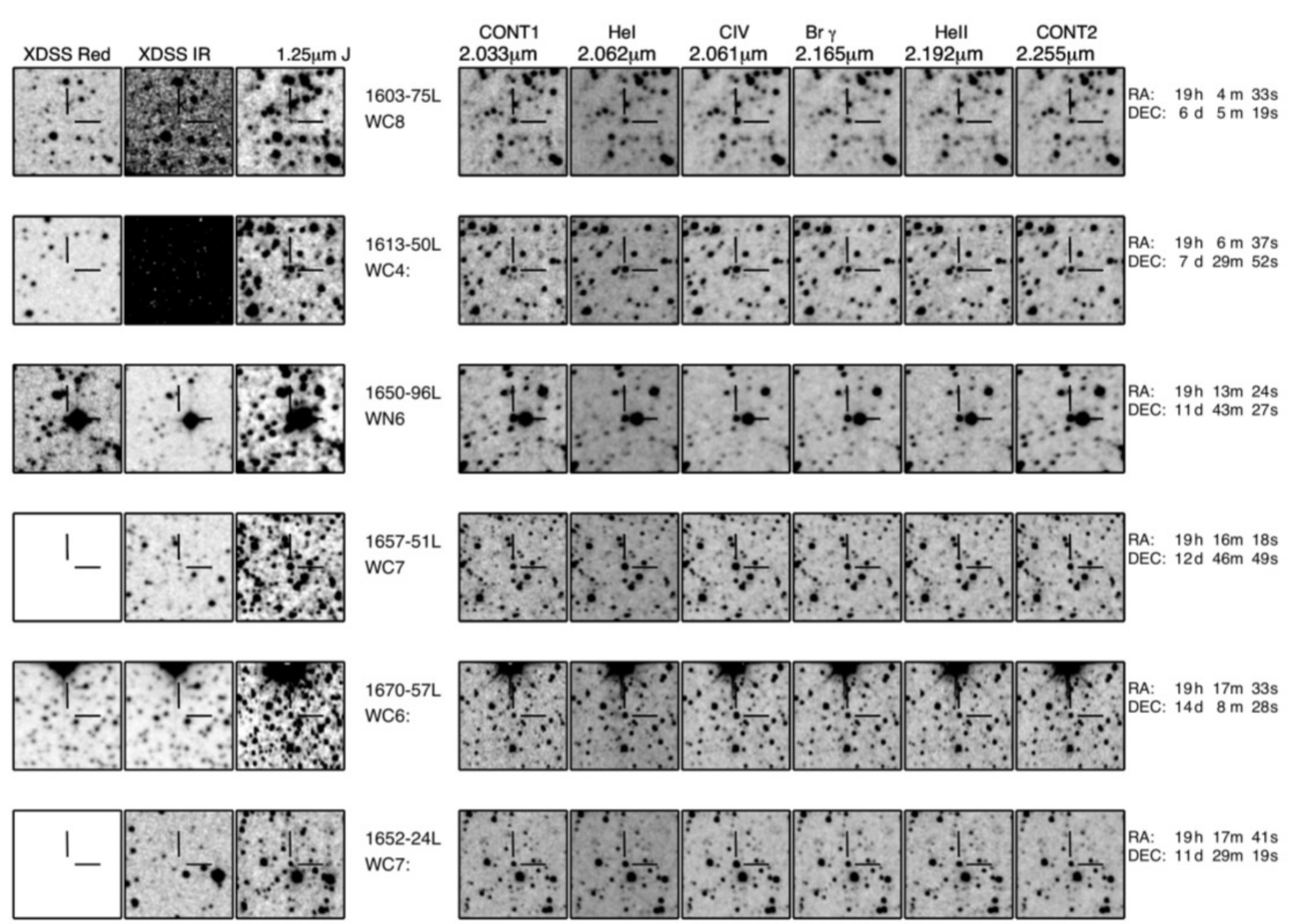}
\end{centering}
\caption{Finder charts for WC and WN stars observed with SpeX.} 
\label{fig:finder2}
\end{figure*}

\begin{figure*}[htbp]
\begin{centering}
\epsscale{.8}
\includegraphics[width=1.1\hsize,angle=90]{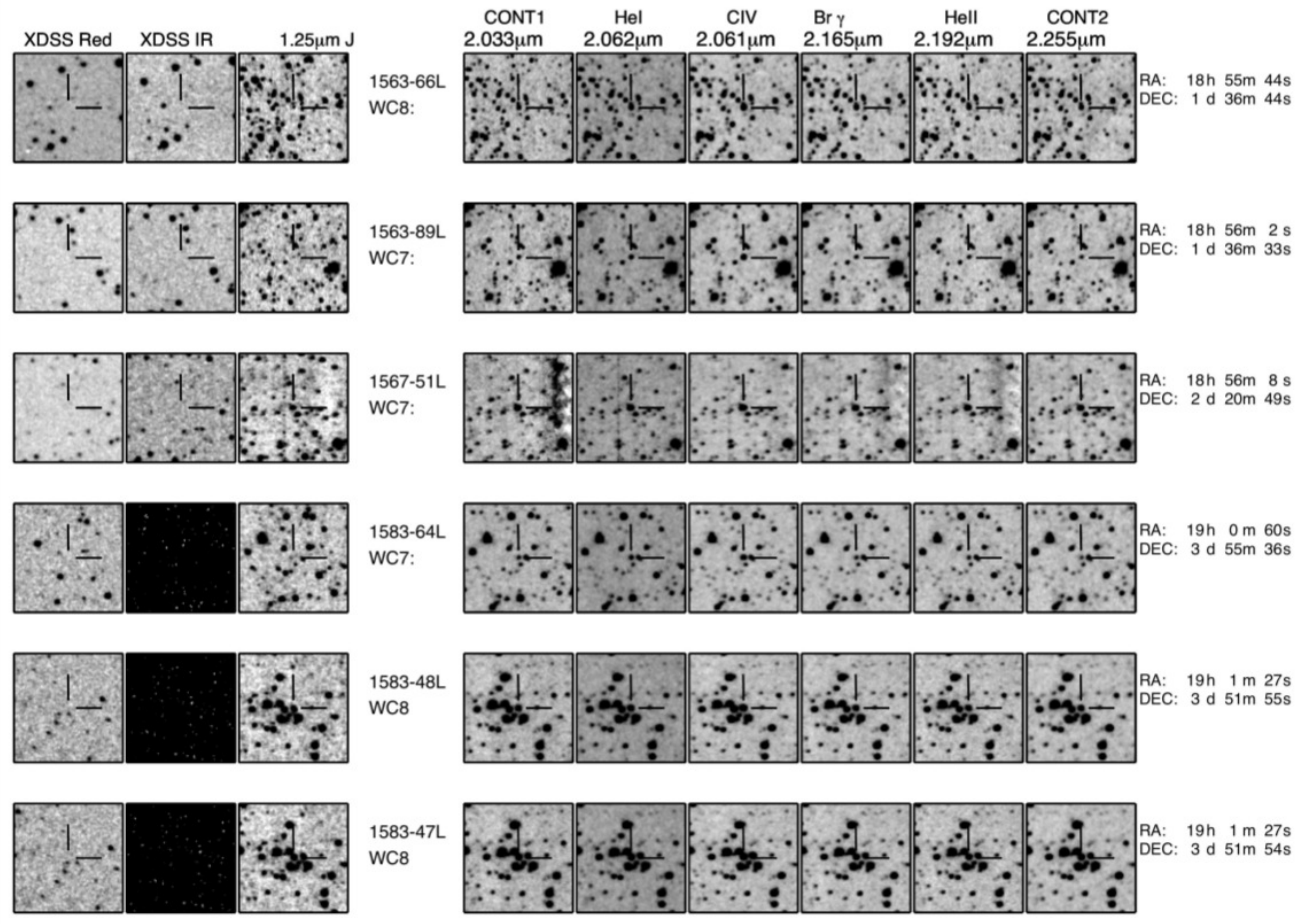}
\end{centering}
\caption{Finder charts for WC and WN stars observed with SpeX.} 
\label{fig:finder3}
\end{figure*}

\begin{figure*}[htbp]
\begin{centering}
\epsscale{.8}
\includegraphics[width=1.1\hsize,angle=90]{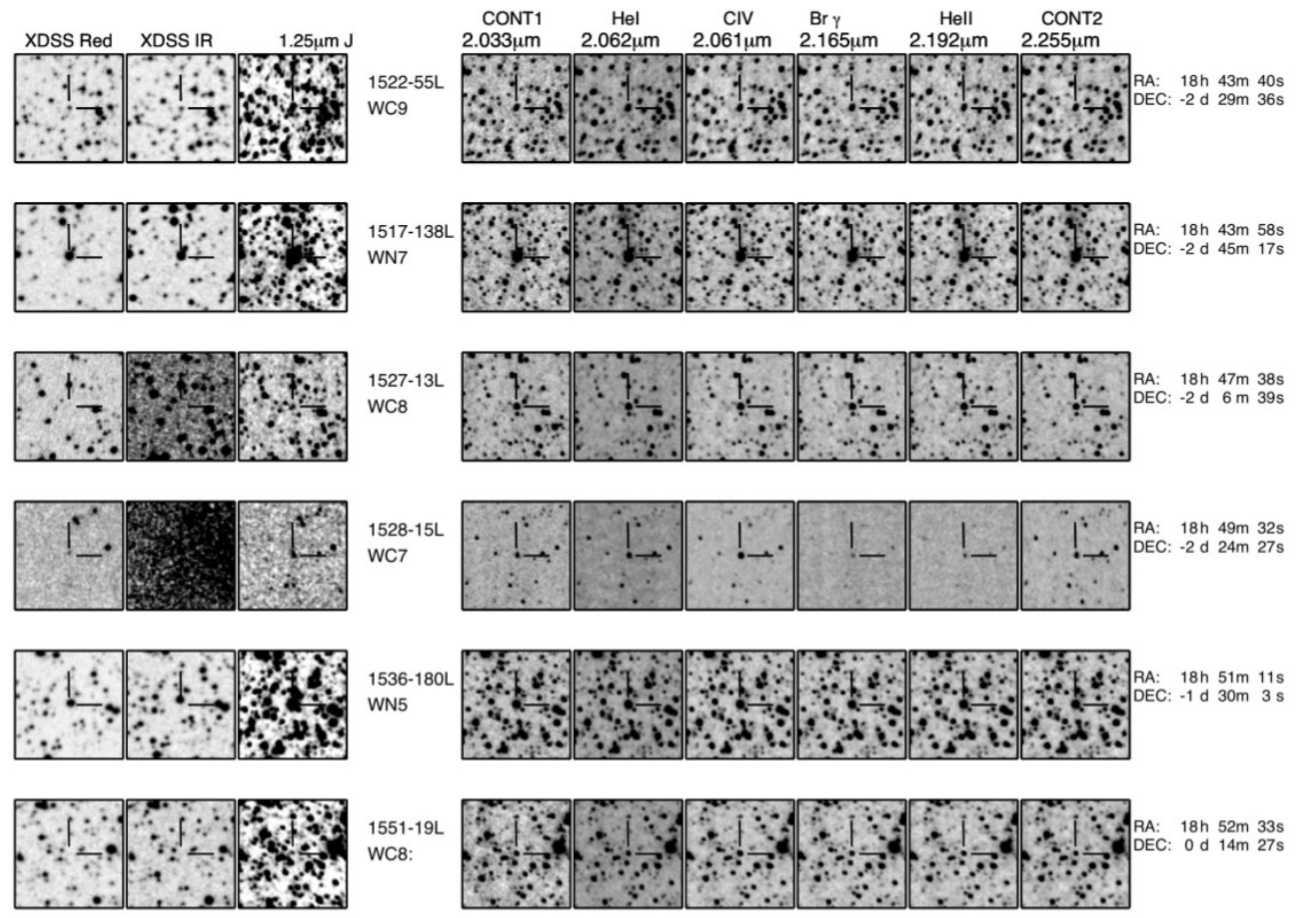}
\end{centering}
\caption{Finder charts for WC and WN stars observed with SpeX.} 
\label{fig:finder4}
\end{figure*}

\begin{figure*}[htbp]
\begin{centering}
\epsscale{.8}
\includegraphics[width=1.1\hsize,angle=90]{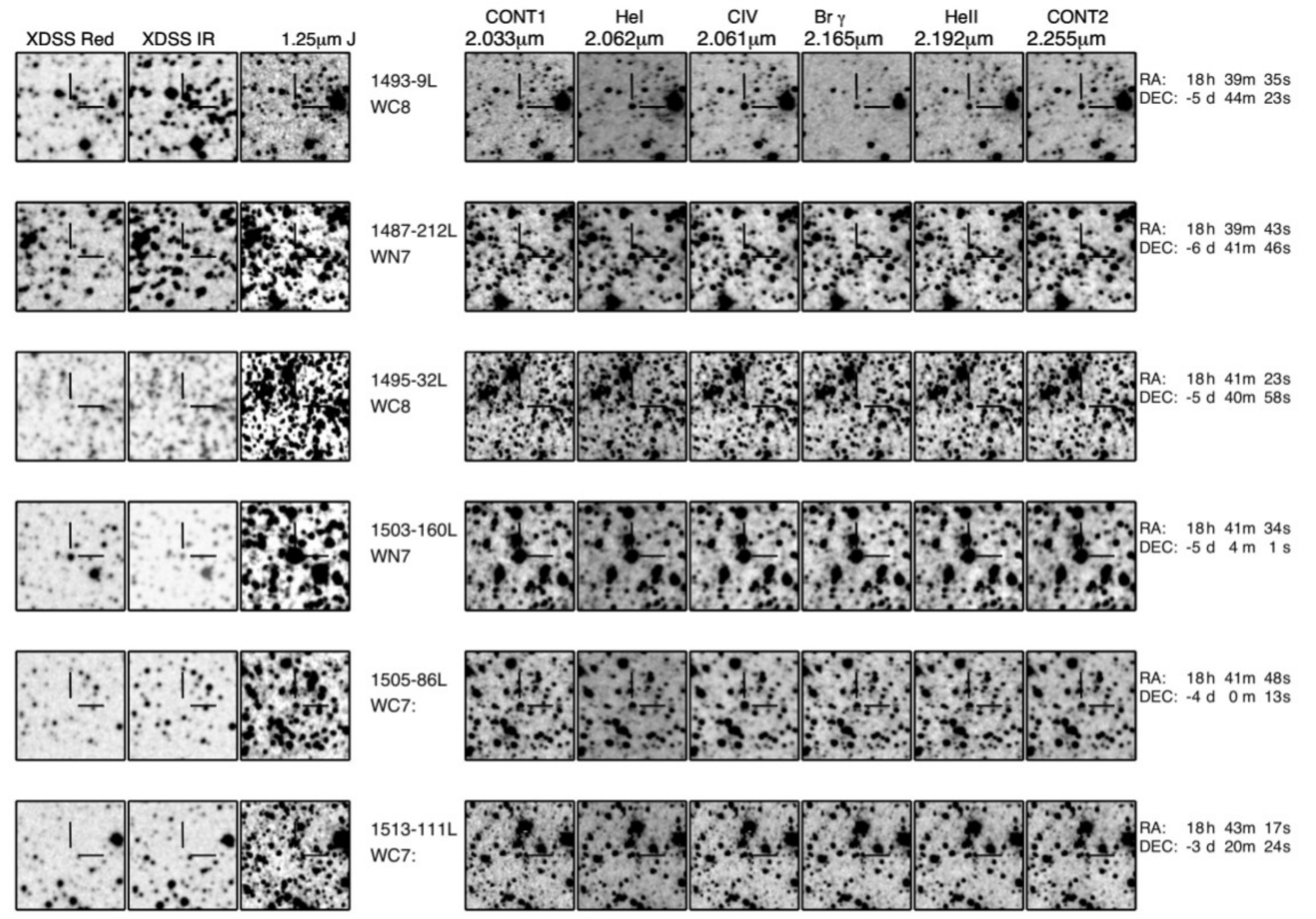}
\end{centering}
\caption{Finder charts for WC and WN stars observed with SpeX.} 
\label{fig:finder5}
\end{figure*}

\begin{figure*}[htbp]
\begin{centering}
\epsscale{.8}
\includegraphics[width=1.1\hsize,angle=90]{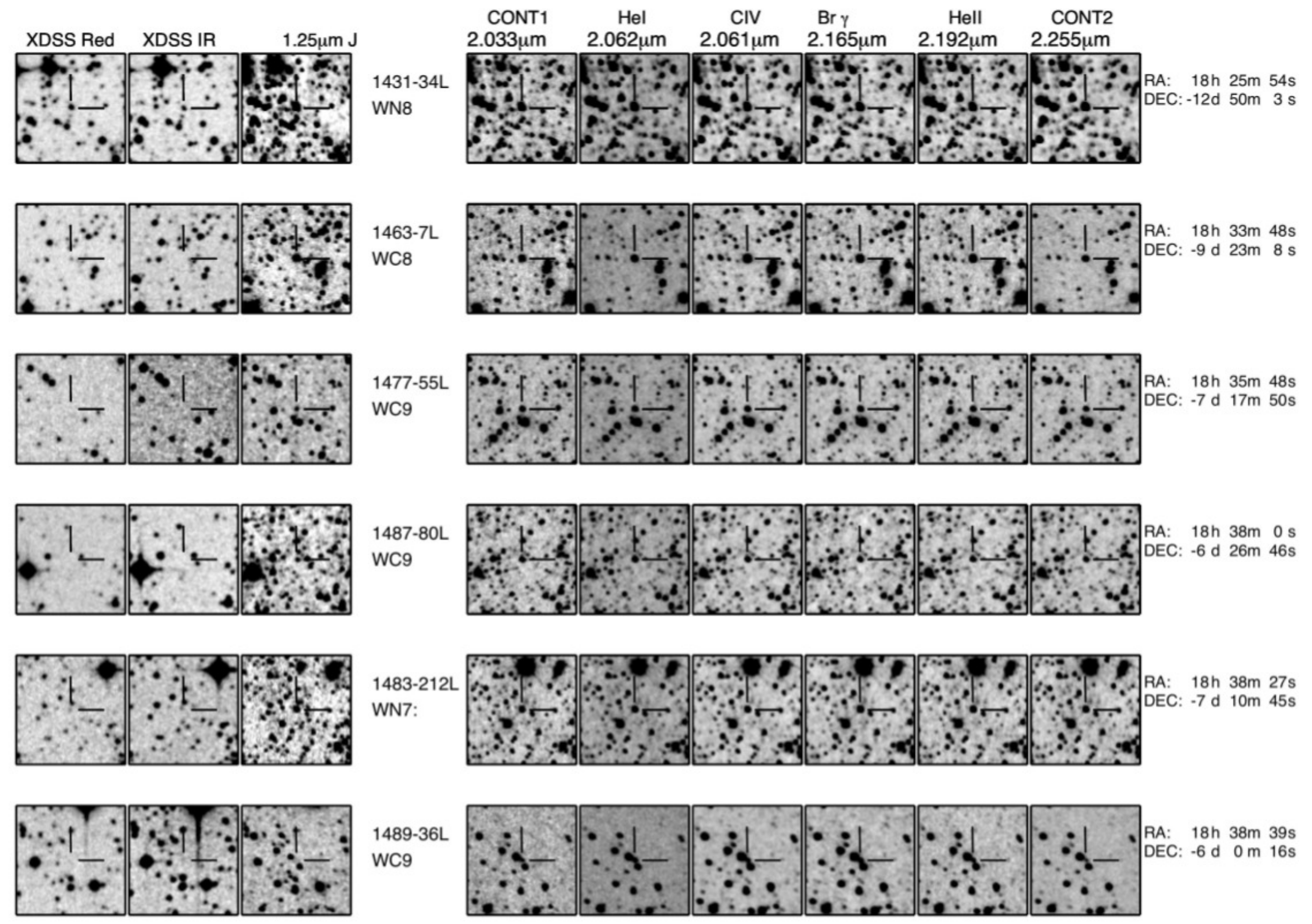}
\end{centering}
\caption{Finder charts for WC and WN stars observed with SpeX.} 
\label{fig:finder6}
\end{figure*}

\begin{figure*}[htbp]
\begin{centering}
\epsscale{.8}
\includegraphics[width=1.1\hsize,angle=90]{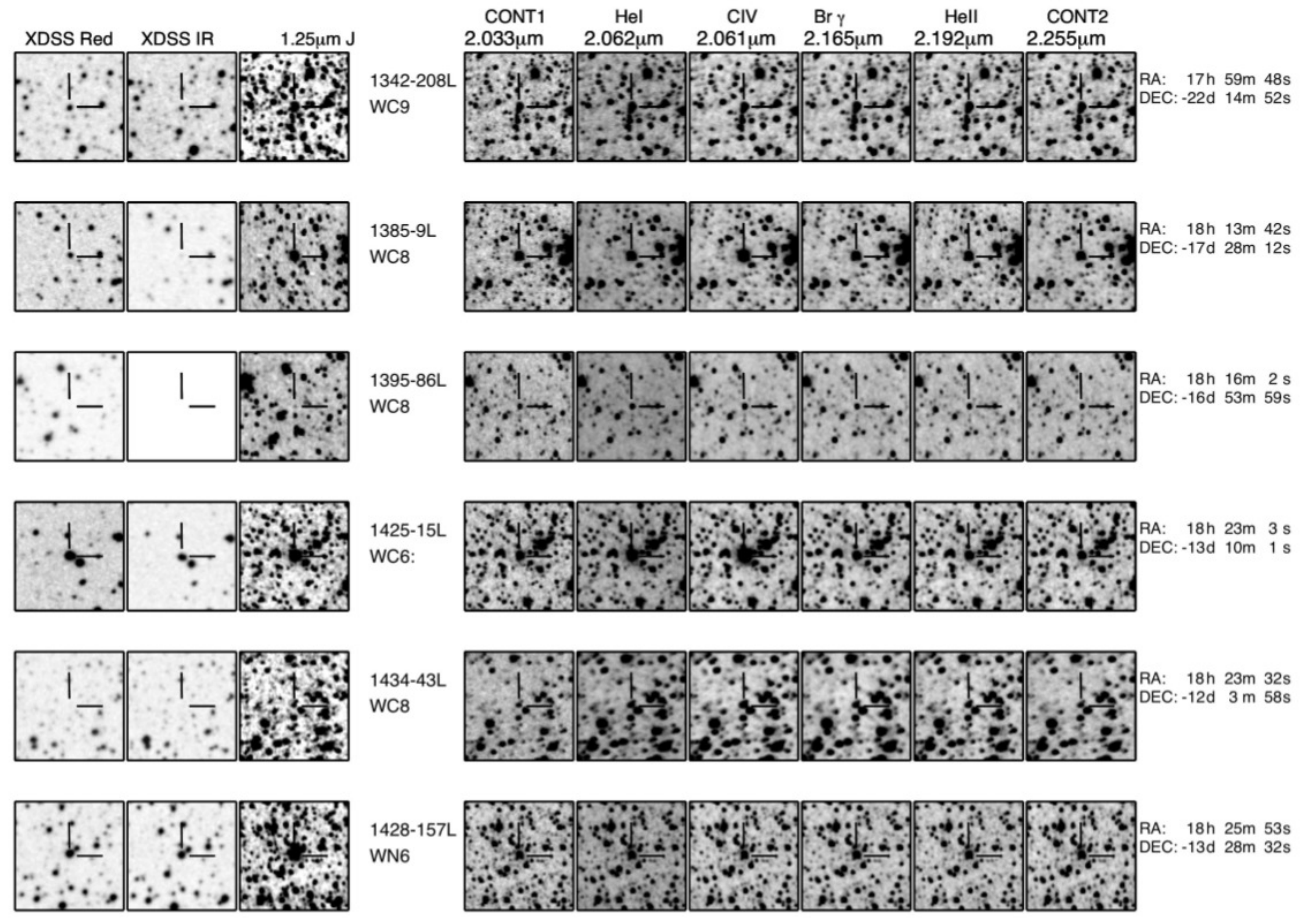}
\end{centering}
\caption{Finder charts for WC and WN stars observed with SpeX.} 
\label{fig:finder7}
\end{figure*}

\begin{figure*}[htbp]
\begin{centering}
\epsscale{.8}
\includegraphics[width=1.1\hsize,angle=90]{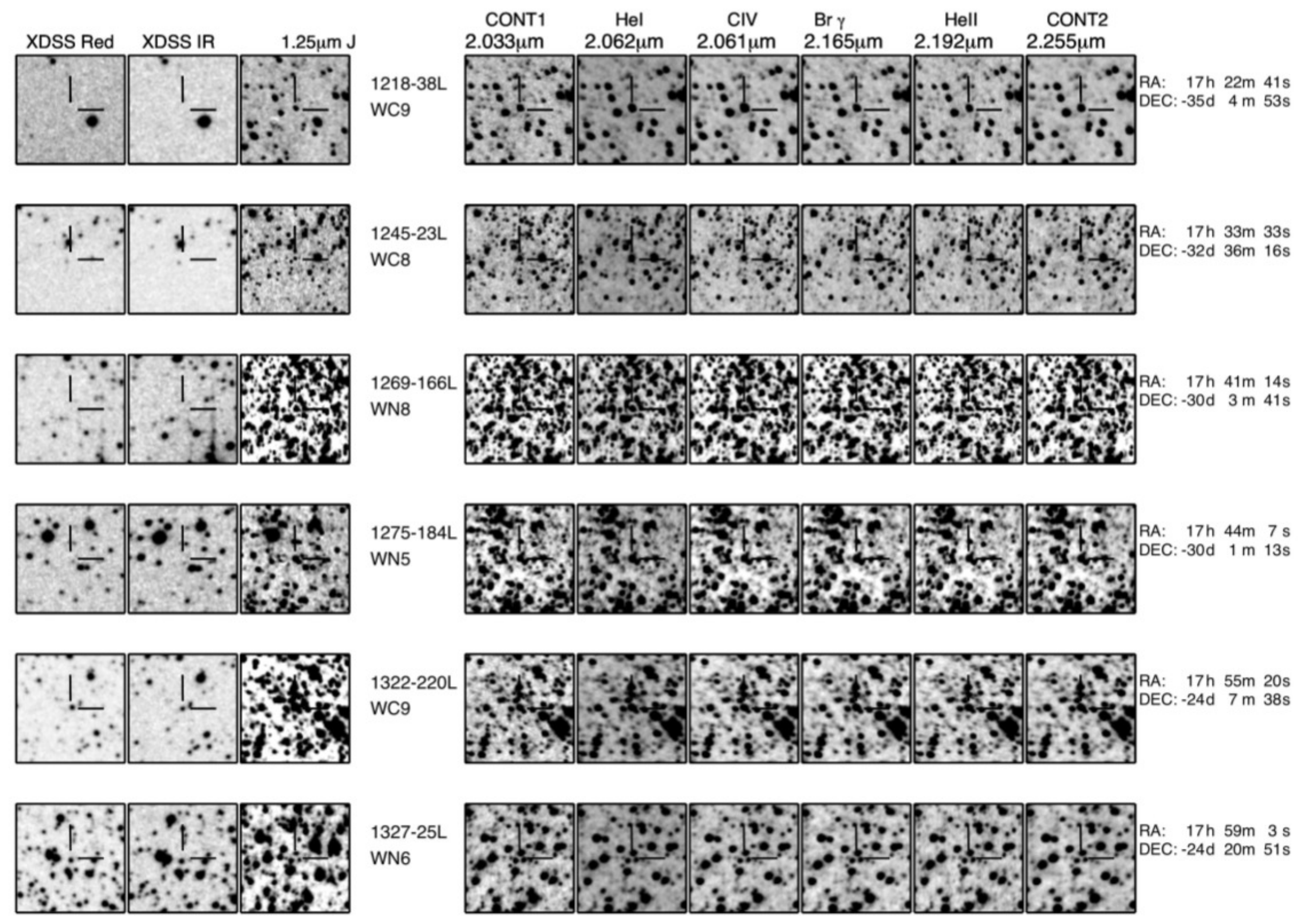}
\end{centering}
\caption{Finder charts for WC and WN stars observed with SpeX.} 
\label{fig:finder8}
\end{figure*}

\begin{figure*}[htbp]
\begin{centering}
\epsscale{.8}
\includegraphics[width=1.1\hsize,angle=90]{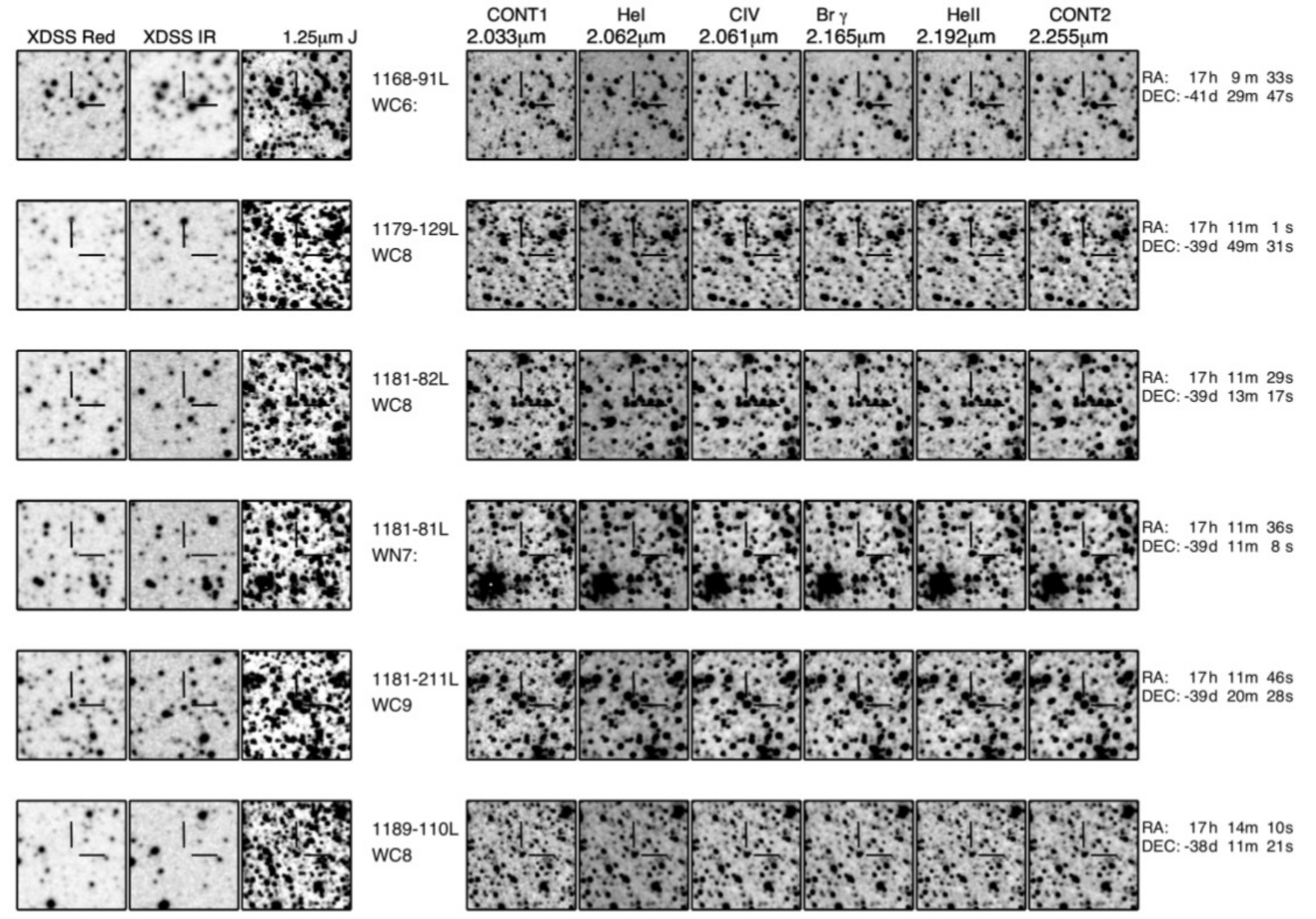}
\end{centering}
\caption{Finder charts for WC and WN stars observed with SpeX.} 
\label{fig:finder9}
\end{figure*}

\begin{figure*}[htbp]
\begin{centering}
\epsscale{.8}
\includegraphics[width=1.1\hsize,angle=90]{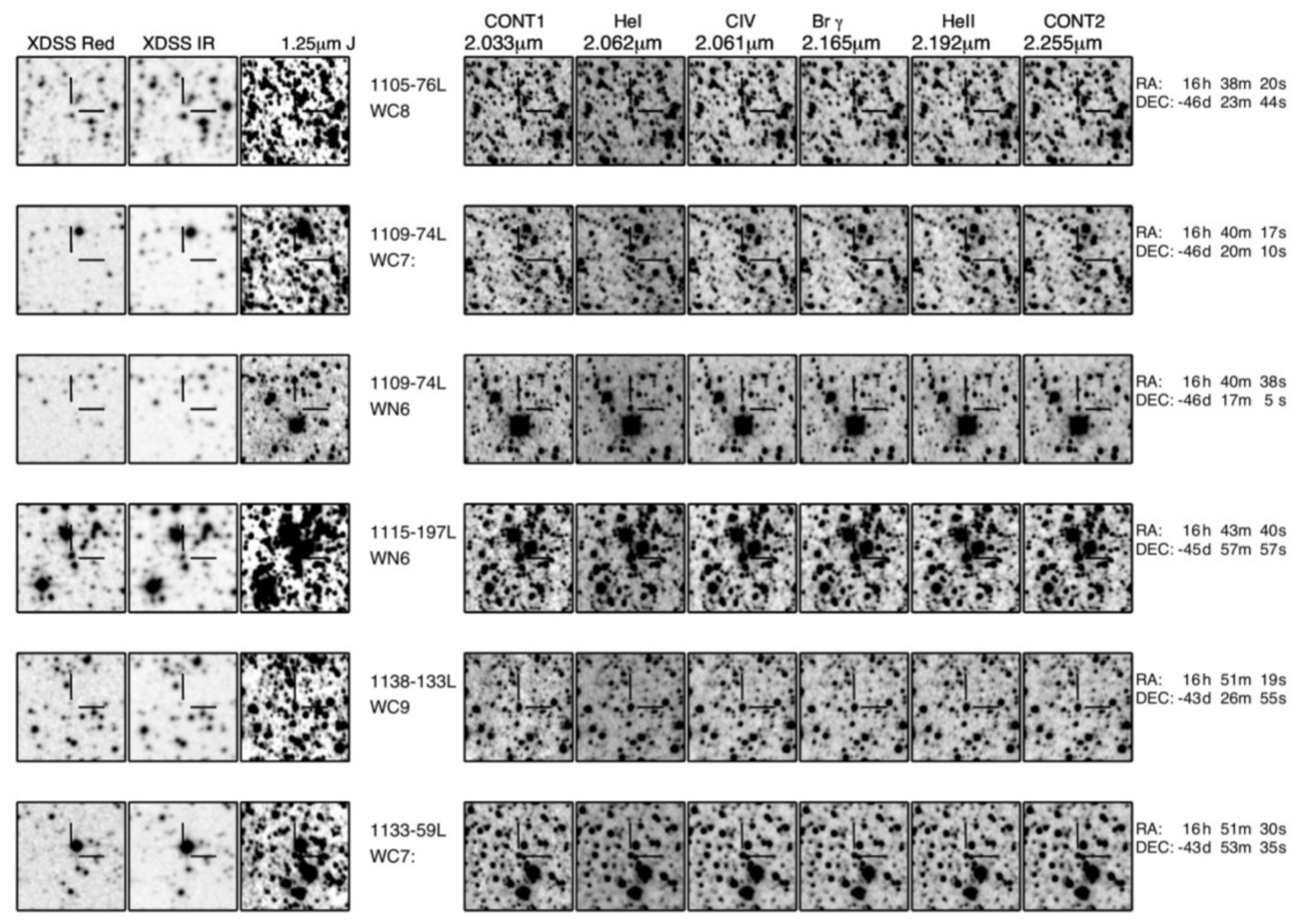}
\end{centering}
\caption{Finder charts for WC and WN stars observed with SpeX.} 
\label{fig:finder10}
\end{figure*}

\begin{figure*}[htbp]
\begin{centering}
\epsscale{.8}
\includegraphics[width=1.1\hsize,angle=90]{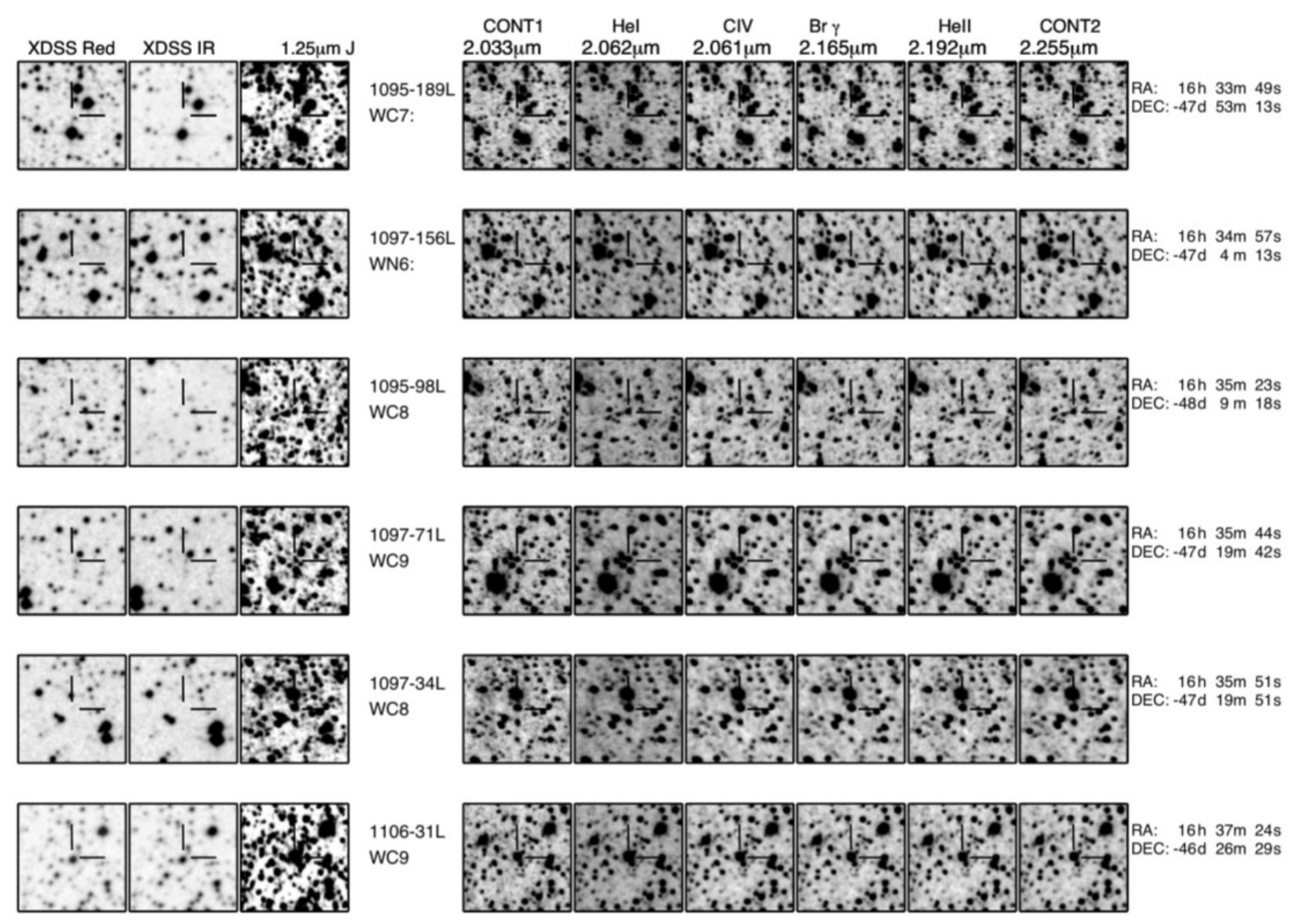}
\end{centering}
\caption{Finder charts for WC and WN stars observed with SpeX.} 
\label{fig:finder11}
\end{figure*}

\begin{figure*}[htbp]
\begin{centering}
\epsscale{.8}
\includegraphics[width=1.1\hsize,angle=90]{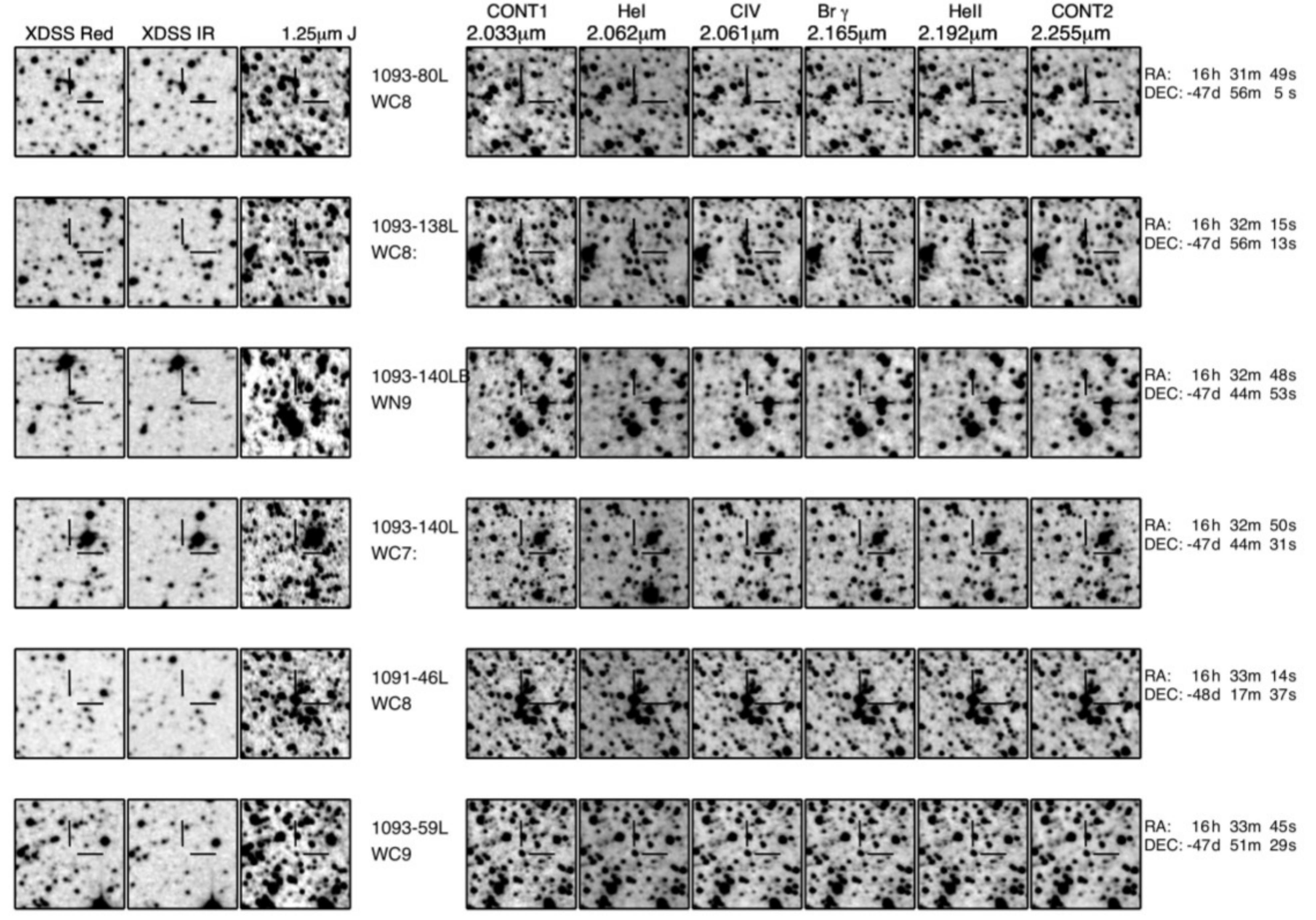}
\end{centering}
\caption{Finder charts for WC and WN stars observed with SpeX.} 
\label{fig:finder12}
\end{figure*}

\begin{figure*}[htbp]
\begin{centering}
\epsscale{.8}
\includegraphics[width=1.1\hsize,angle=90]{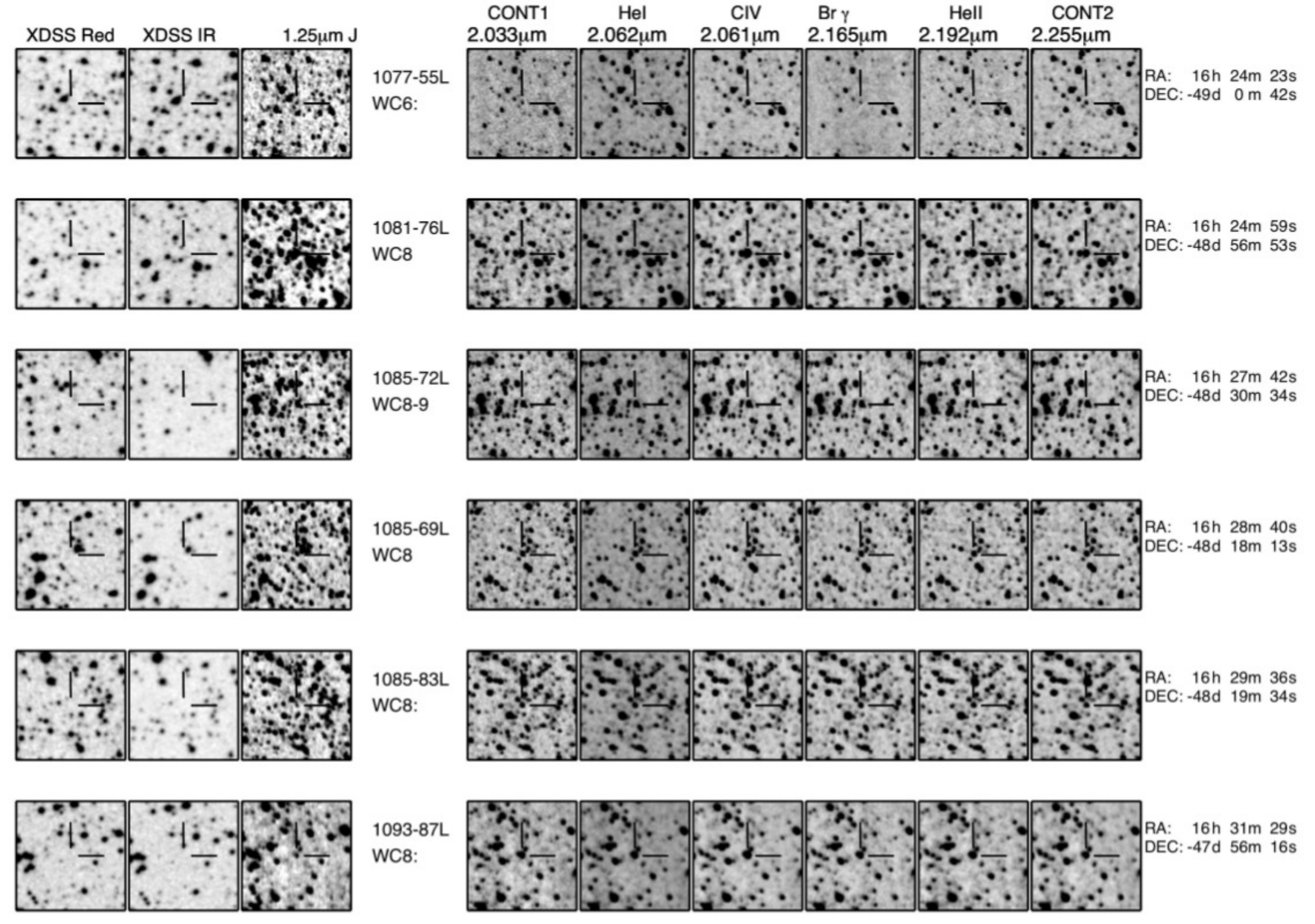}
\end{centering}
\caption{Finder charts for WC and WN stars observed with SpeX.} 
\label{fig:finder13}
\end{figure*}

\begin{figure*}[htbp]
\begin{centering}
\epsscale{.8}
\includegraphics[width=1.1\hsize,angle=90]{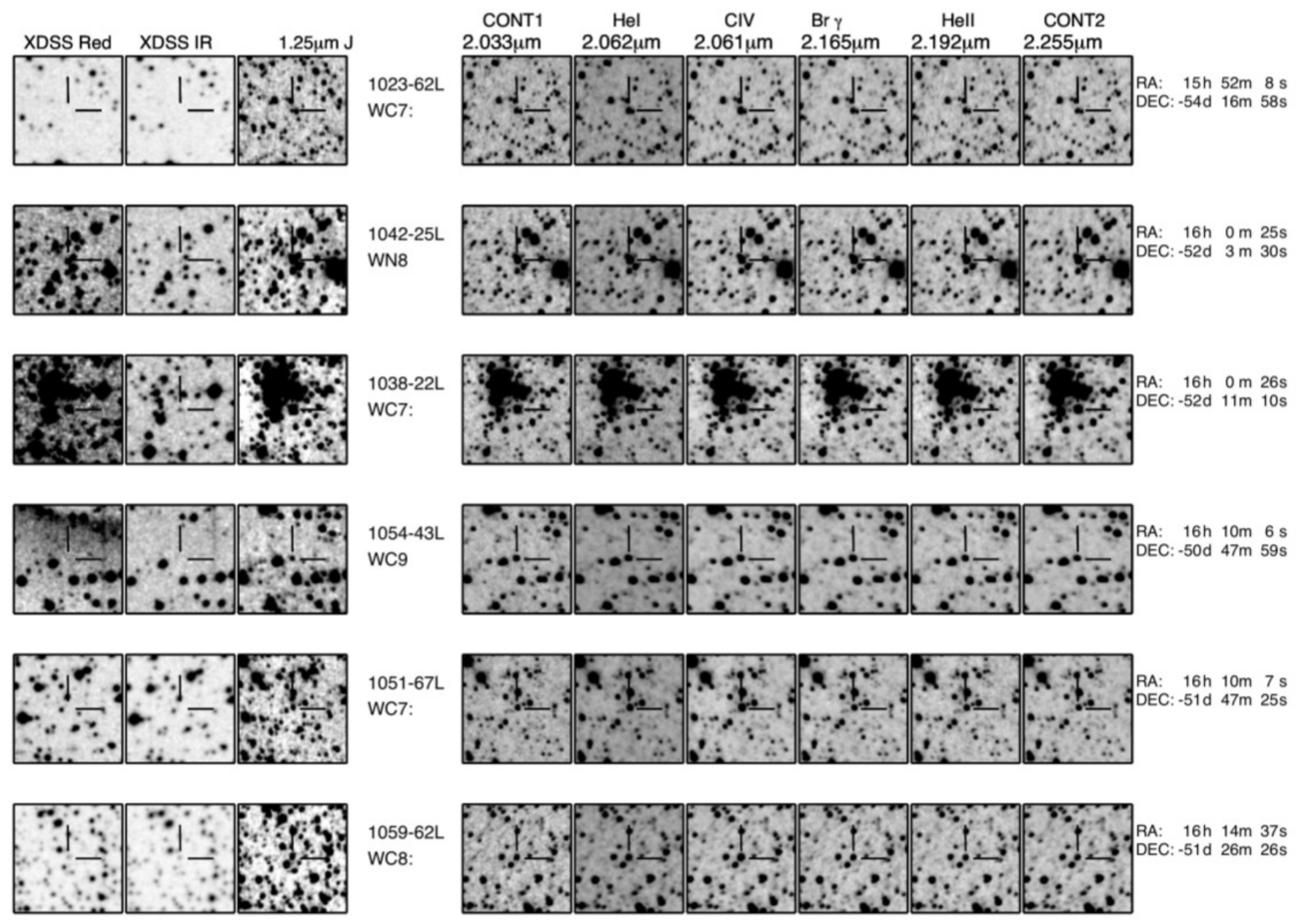}
\end{centering}
\caption{Finder charts for WC and WN stars observed with SpeX.} 
\label{fig:finder14}
\end{figure*}

We present in Figures~\ref{fig:finder1}-~\ref{fig:finder14} the finder charts for the 71 new Wolf-Rayet stars as well as the 11 previously identified objects described in this paper.

\clearpage
\pagebreak

\clearpage

\section{Conclusions}\label{conclusion}
We have discovered 71 new Galactic WR stars, 17 of type WN and 54 of type WC, via our near-infrared narrow-band survey of the Galactic plane. The reduced extinction from dust and gas in the near infrared makes this a highly effective method for future discovery of the thousands of undetected Galactic WR stars. Of the 146 total candidates observed spectrographically, 83 proved to be new or previously identified WR stars.  With such a 57\% detection rate, we have barely scratched the surface of the wealth of new WR stars expected to be discovered within our survey area with the available data.

An initially fairly simple sky-subtraction methodology (used in Paper I) resulted in relatively scattered color-magnitude diagrams,
and a detection efficiency of 24\%. By raising our cut for emission objects in the study reported here to 5$\sigma$ we have also increased our detection efficiency to 57\%. Most of our non-detections were erroneously selected objects with almost featureless spectra, and absorption bands in our continuum filters that mimicked emission lines. Improved sky subtraction (using weeks of data, median-filtered in each filter as skyflats) and including J,H,K and mid-IR photometry of our candidates (the complementary method of \citep{mau11}) will allow us to further improve the detection rate of emission-line objects. We expect this survey to yield thousands of additional WR star discoveries in the coming years.  

Our survey limits will be pushed fainter by the use of still larger infrared telescopes for spectroscopic follow-up. As we increase the number of known stars, we will also increase the statistical significance of distribution plots, and subtype abundances, allowing us to learn more about our Galaxy's structure and composition. The Galactic center is expected to prove an especially rich area for discovery, but it is still largely terra incognita as the crowding of stars there is very high. The large majority of Galactic Wolf-Rayet stars remain to be discovered, but we now have a proven and highly efficient technique to greatly extend the search.

\acknowledgments
MMS, JF, JG,  and DZ acknowledge with gratitude Hilary and Ethel Lipsitz, whose ongoing support has been essential to the success of this program. AFJM, RD, LD are grateful to NSERC (Canada) and FQRNT (Quebec) for financial aid. We also thank the American Museum of Natural History for essential funding.

\clearpage
\bibliographystyle{apj}
\bibliography{paper}

\end{document}